\documentclass[twocolumn, showpacs, aps, pre, superscriptaddress, amsmath, amssymb, nofootinbib]{revtex4-1}

\usepackage{xcolor}
\usepackage{float}
\usepackage[utf8]{inputenc}
\usepackage{graphicx}
\usepackage{amsmath}
\usepackage[linktocpage,colorlinks=true,allcolors=blue,breaklinks=true]{hyperref}
\usepackage{ulem}
\usepackage{units}
\usepackage{upgreek}
\usepackage[title]{appendix}
\usepackage{enumitem}
\usepackage{mathtools}
\usepackage{booktabs}
\usepackage{multirow}
\usepackage{array}
\usepackage[caption=false]{subfig}

\makeatletter
\renewcommand{\p@subsection}{}
\makeatother

\newcommand{\Da}{D^\mathrm{a}_i}
\newcommand{\D}{D^\mathrm{a}}
\newcommand{\cc }{c^\mathrm{a}}
\newcommand{\cb}{\bar{c}^\mathrm{a}}
\newcommand{\ca}{c^\mathrm{a}_i}
\newcommand{\cab}{\bar{c}^\mathrm{a}_i}
\newcommand{\cat}{\tilde{c}^\mathrm{a}_i}

\newcommand{\ci}{\boldsymbol{c}}
\newcommand{\cin}{c^\mathrm{0}_i}
\newcommand{\cm}{c^\mathrm{m}_i}
\newcommand{\bcm}{\bar{c}^\mathrm{m}_i}
\newcommand{\ccm}{c^\mathrm{m}}
\newcommand{\cmb}{\bar{c}^\mathrm{m}}
\newcommand{\h}{h}
\newcommand{\Dmu}{D^\mathrm{m}}
\newcommand{\Dm}{D^\mathrm{m}_i}
\newcommand{\lambdai}{\lambda_i}
\newcommand{\g}{\gamma^\mathrm{a}}
\newcommand{\ga}{\gamma^\mathrm{a}_i}
\newcommand{\gm}{\gamma^\mathrm{m}_i}
\newcommand{\alphai}{\alpha_i}
\newcommand{\Kpp}{K^\mathrm{p}}
\newcommand{\Kp}{K^\mathrm{p}_i}
\newcommand{\Ka}{K^\mathrm{a}_i}
\newcommand{\Ej}{e_j}

\newcommand{\Aji}{A_{ji}}
\newcommand{\Sji}{S_{ji}}
\newcommand{\aii}{A_{ii}}

\newcommand{\SigmaC}{\Sigma^\mathrm{c}}
\newcommand{\No}{{N_\mathrm{o}}}
\newcommand{\Nr}{{N_\mathrm{r}}}
\newcommand{\Penv}{P_\mathrm{env}}
\newcommand{\norm}{\mathcal{N}}
\newcommand{\nMax}{n_\mathrm{max}}

\newcommand{\diff}{\mathrm{d}}
\newcommand{\vect}[1]{\boldsymbol{#1}}
\newcommand{\tens}[1]{\boldsymbol{#1}}
\newcommand{\Eqref}[1]{Eq.~\eqref{#1}}

\begin{document}

\title{Zonal receptor distributions maximize olfactory information}

\author{Swati Sen}
\affiliation{Max Planck Institute for Dynamics and Self-Organization, Am Fassberg 17, 37077 G{\"o}ttingen, Germany}

\author{David Zwicker}
\email{david.zwicker@ds.mpg.de}
\affiliation{Max Planck Institute for Dynamics and Self-Organization, Am Fassberg 17, 37077 G{\"o}ttingen, Germany}

\date{\today}

\begin{abstract}
{
The olfactory sense measures the chemical composition of the environment using a diverse array of olfactory receptors.
In vertebrates, the olfactory receptors reside in a mucus layer in the nasal cavity and can thus only detect odorants that are inhaled with the airflow and dissolved in mucus.
These physical processes fundamentally affect how many odorant molecules contact the receptors.
We hypothesize that the olfactory system works efficiently by optimizing the placement of receptors for maximal information transmission.
Using a simplified model, we capture all relevant physical processes and show that odorant concentrations generally exhibit an exponential distribution.
Combining this result with information theory, we further show that receptors separated into distinct spatial zones maximize the transmitted information.
Our results are consistent with experimentally observed receptors zones and might help to improve artificial smell sensors.
}

\end{abstract}

\pacs{}

\maketitle

\begin{figure}[t]
    \includegraphics[width=\columnwidth]{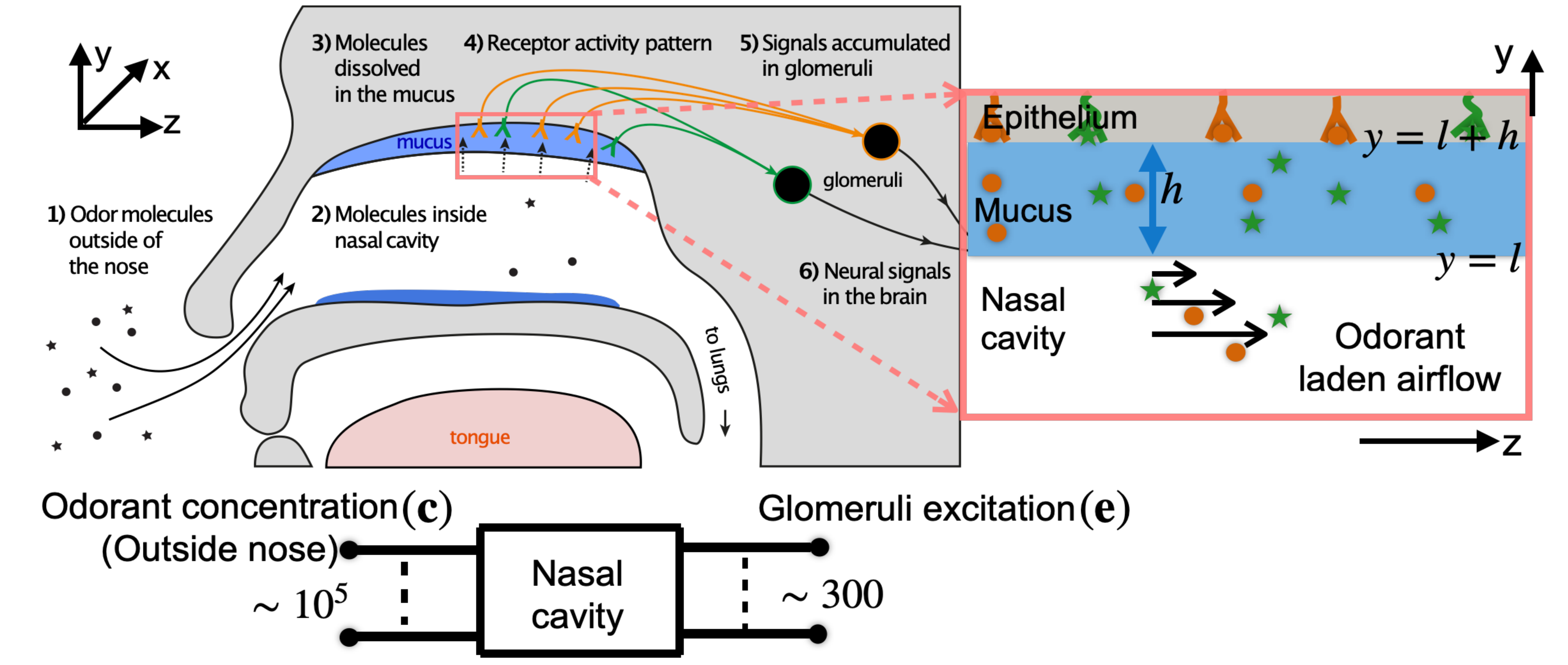}
    \caption{\textbf{Olfactory system as a communication channel.}
    Schematic representation of olfactory processes in the nasal cavity.
    The inset on the right shows the cavity wall comprised of an epithelium covered with a mucus layer of thickness $\h$ in which the odoroant receptors reside. 
    The schematic picture at the bottom depicts the nasal cavity as a communication channel transforming information about the odor (the odorant concentration vector $\vect{c}$) to an excitation pattern of the $\Nr \sim 300$ types of receptors in human.
    }
    \label{fig1:generalpic}
\end{figure}

Olfaction is a crucial sense, which measures the chemical composition of the environment.
Vertebrates generally sample ambient air by sniffing, which brings odorant molecules into the nasal cavity with the inhaled air.
Similarly, mammals transport food odors from the pharynx to the nasal cavity during exhalation in a process called retronasal smell.
In both cases, odorant molecules dissolve in the mucus layer lining the walls of the nasal cavity, where they come in contact with olfactory receptors~\cite{silvachemsense2016}.
The information about the diverse odorants in natural smells~\cite{KnudsenPhytochemistry1993} is encoded by many different types of olfactory receptors.
The response of all receptors of the same type is accumulated in corresponding glomeruli and the combined response is forwarded to the brain, where the information is decoded~\cite{MAINLAND2014, Touhara2009}.
Taken together, the olfactory system can be described as a communication channel, where physical and neural processes transform the ambient odorant concentration profile into neural activity patterns; see Fig.~\ref{fig1:generalpic}.
Since the olfactory system has likely been optimized evolutionarily, information theory can be used to understand its specific design~\cite{Atick2011, Bialek2016}.
Maximizing transmitted information has already been useful to understand neural encoding of olfactory information~\cite{TiberiueLife2019,ZwickerPNAS2016}, but an analysis of the physical processes is still missing.

Physical processes fundamentally affect how many molecules of each odorant reach the olfactory receptors.
It is thus likely that olfactory receptors are placed to optimize odorant sensing.
In fact, experiments have shown that olfactory receptors are not homogeneously distributed in the nasal cavity, but rather separate into different zones~\cite{tanchemsense2018,Horowitz12241,vedinJCN2009,AxelCell1993,Resslercell1993}.
We hypothesize that such a zonal distribution optimizes sensing of physically distinct odorants to maximize information transmission.
To test this hypothesis, we here develop a simplified model of the olfactory processing, which incorporates advective transport, adsorption in mucus, and detection by heterogeneously distributed receptors.
In contrast to detailed models based on computational fluid dynamics~\cite{Barbarite2021, bruning2020, Chengyu2018}, our analytical results enable an analysis of the downstream processing.
Using information theory, we then show that a zonal distribution indeed maximizes the information the brain receives about the ambient odor.

\section{Results}
The paper is separated into two parts:
First, we analyze the physical processes of odor advection and dissolution to predict the odorant concentration profile at the receptors in sections \ref{sec:airflow}--\ref{sec:mucus}.
We then use these results in combination with information theory to predict optimal receptor distributions in sections \ref{sec:excitation}--\ref{sec:zonal_distribution}.

\subsection{Airflow inside nasal cavity is laminar}
\label{sec:airflow}

The nasal cavity has a rigid and narrow geometry, which suppresses turbulent flow effectively~\cite{DavidPNAS2018}.
Moreover, the direction of flow changes only little along the length~$L$ of the cavity, so that it locally is well approximated by a parabolic flow between two parallel cavity walls separated by a short distance~$2l$ ($l \ll L$).
To describe the odor transport in the cavity, we can thus orient the $z$-axis along the main direction of flow from anterior to posterior and we choose the $y$-axis such that it is (locally) perpendicular to the walls; see Fig.~\ref{fig1:generalpic}.
The flow and the associated odor transport is invariant in the perpendicular $x$-direction, so the  problem reduces to two dimensions.
During restful breathing, the flow can also be approximated as stationary and incompressible~\cite{DavidPNAS2018}, implying the velocity is oriented in the $z$-direction and its magnitude is given by
\begin{equation}
    u_z = \frac{3u_0}{2l^2}\left(l^2-y^2\right) 
    \;,
    \label{eqn:flow_speed}
\end{equation} 
where $y\in[-l,l]$ and $u_0$ is the mean instantaneous velocity; see Fig.~\ref{fig1:generalpic}.
Typical values of the parameters for humans are summarized in Table \ref{table1}.
Note that olfactory processing is generally synchronized with sniffing~\cite{smear2011, shusterman2011}, implying that the temporal details of the airflow are less important.

\begin{table}
    \centering
    \caption{Typical parameter values for human nasal cavities}
    \begin{tabular}{lll}
        \toprule[1pt]
        \addlinespace[2pt]
        Quantity & Symbol & Value \\
        \addlinespace[2pt]
        \midrule[1pt]
        \addlinespace[2pt]
        Cavity length~\cite{DavidPNAS2018} & $L$ & \unit[10]{cm} \\
        Air mean speed~\cite{DavidPNAS2018} & $u_0$ & \unitfrac[$10-100$]{cm}{s} \\
        Cavity width~\cite{DavidPNAS2018} & $l$ & \unit[0.3]{cm} \\
        Mucus thickness~\cite{Craven2012} & $\h$ & \unit[$10^{-3}$]{cm}\\
       Diffusivity in air~\cite{Kurtzchemse2004, KEYHANI1997} & $\D$  & \unitfrac[$10^{-2}-0.1$] {cm$^2$}{s} \\
       Diffusivity in mucus~\cite{Craven2012, Kurtzchemse2004} & $\Dmu$ & \unitfrac[$ 10^{-6}- 10^{-5}$]{cm$^2$}{s} \\
       Péclet number~\cite{Tritton} & $Pe = \frac{u_0L}{\D}$ & \unit[$10^{3}-10^{4}$] \\
       Mucus velocity~\cite{shangPLoSOne2021} & $u^\mathrm{m}$ & \unitfrac[$10^{-3}-10^{-2}$] {cm}{s} \\
        Adsorption coefficient~\cite{Craven2012} & $\g$ & \unitfrac[$10^{-2}-10^3$]{cm}{s} \\
        Solubility in mucus~\cite{Craven2012, Kurtzchemse2004}& $\Kpp$ & \unit[$10^{-6}-1$] \\
        \addlinespace[2pt]
        \bottomrule[1pt]
    \end{tabular}
    \label{table1}
\end{table}

\subsection{Odorant concentration decays exponentially along nasal cavity}
\begin{figure}[t]
    \centering
    \includegraphics[scale=0.16]{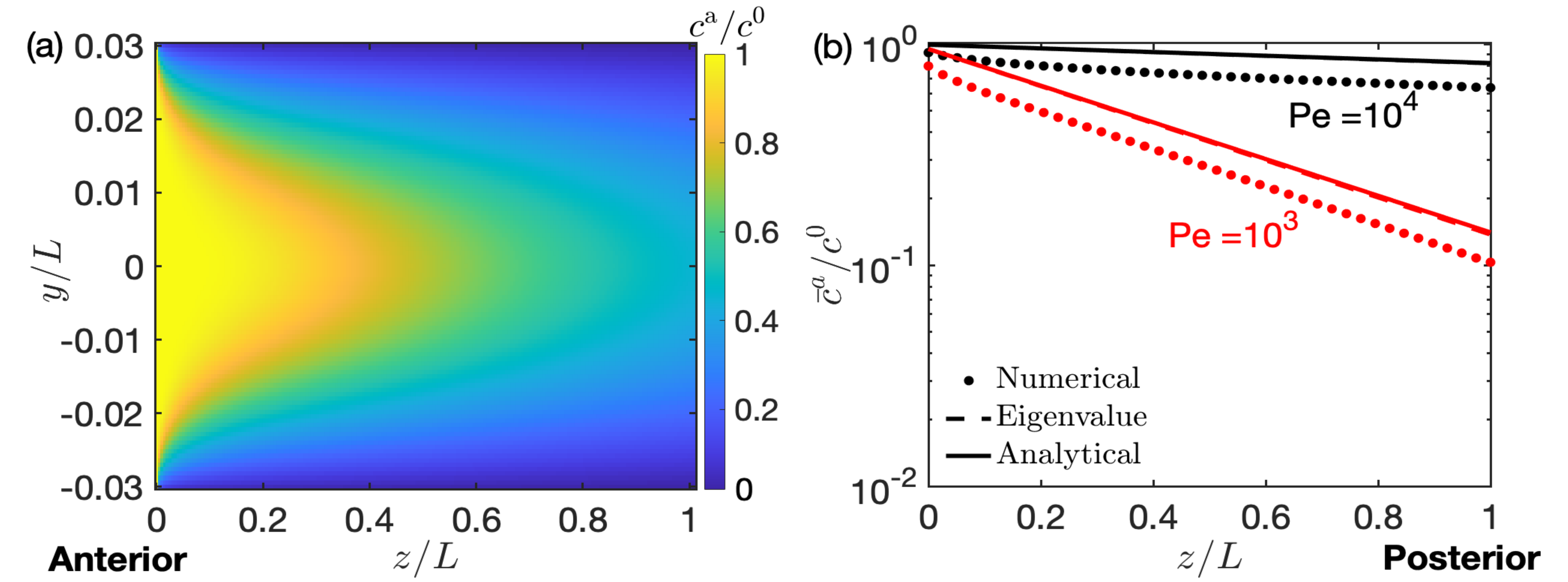}
    \caption{\textbf{Odorant concentration decreases exponentially in the nasal cavity.}
    (a) Odorant concentration $\cc$ inside the nasal cavity as a function of distances~$y$ and $z$ across and along the cavity, respectively.
    Data has been obtained from numerical simulations of Eq.~\eqref{eqn:advection_diffusion_stationary}  for $\mathrm{Pe}=10^3$.
    (b) Cross-sectionally averaged concentration $\cb$ as a function of $z$ for $\mathrm{Pe} \in \{10^3, 10^4\}$.
    The analytical approximation (solid lines) predicts the exponential decay obtained from numerical simulations (dots) and the eigenvalue analysis (dashed lines).
    (a, b) Model parameters are $L=\unit[10]{cm}$, $l=\unit[0.3]{cm}$, $\D=\unitfrac[0.1]{cm^2}{s}$ and $\g=\unitfrac[10]{cm}{s}$.
    }
    \label{fig2:odorincavity}
\end{figure}

We next consider how odors are transported with the airflow.
For simplicity, we focus on orthonasal smell, i.e., how odorants are transported from the ambient air into the cavity during inhalation.
However, the converse case of retronasal smell can be treated analogously.
Since odors typically comprise many different kinds of odorant molecules, we seek the concentration $\ca$ of the $i$-th odorant in the nasal cavity, which is governed by an advection-diffusion equation,
\begin{equation}
  \partial_t \ca = \Da \nabla^2\ca - u_z \frac{\partial \ca}{\partial z}
  \;,
  \label{eqn:advection_diffusion}
\end{equation}
where $\Da$ is the diffusivity in air and $u_z$ is given by Eq.~\eqref{eqn:flow_speed}.
The adsorption of odors at the walls of the cavity is described by the boundary condition
\begin{align}
  \label{eqn:advection_diffusion_bc}
    \pm \Da \partial_y \ca &= \ga\ca
    & \text{at} &&
    y&=\mp l
    \;,
\end{align}
where $\ga$ quantifies the adsorption of odorant~$i$, which we discuss in more detail below.
Additionally, we consider a fixed ambient concentration of $\ca(y,z=0)=\cin$ at the inlet and we assume that advection dominates diffusion at the end of the cavity at $z=L$, implying $\partial_z \ca|_{z=L}=0$.
Similar to the airflow discussed above, we consider the stationary state of Eq.~\eqref{eqn:advection_diffusion}, implying
\begin{equation}
  \Da \left(\partial_y^2\ca + \partial_z^2\ca\right) = u_z \partial_z \ca
  \;,
  \label{eqn:advection_diffusion_stationary}
\end{equation}
where we assumed that axial diffusion is negligible compared to advection ($ \Da \ll u_0 L$).
Fig.~\ref{fig2:odorincavity}a shows numerical solutions of this equation, which demonstrate a strong, nearly exponential variation along the length of the cavity, while the cross-sectional differences are small.
We thus determine the cross-sectionally averaged concentration, $\cab(z) = (2l)^{-1}\int_{-l}^l \ca (y,z) \diff y$, which demonstrates the near-exponential dependence on $z$; see Fig. \ref{fig2:odorincavity}b.

The exponential decay of the odorant concentration along the cavity can also be demonstrated analytically.
To do this, we split off the cross-sectionally averaged part, $\ca(y,z)=\cat(y)\cab(z)$, to use separation of variables on Eq.~\eqref{eqn:advection_diffusion_stationary}.
This analysis implies a mode decomposition of the solutions, where the mode with the longest decay length~$\lambda_i$ dominates the result away from the inlet~\cite{DavidPNAS2018}.
Consequently, we directly find $\cab(z)=c^0_i e^{-z/\lambda_i}$, where $\lambda_i$ is the largest solution to the eigenvalue problem
\begin{equation}
   \frac{\diff^2 \cat(y)}{\diff y^2} = -\lambda_i^{-1} \frac{u_z}{\Da}\cat(y)
   \;.
   \label{eqn:ad_eigenvalues}
\end{equation}
We solve Eq.~\eqref{eqn:ad_eigenvalues} numerically using the appropriate boundary condition $\pm \Da \partial_y \cat = \ga\cat$ at $y=\mp l$.
This leads to an analytical prediction of the decay length scale~$\lambda_i$, which matches the numerical data very well; see Fig.~\ref{fig2:odorincavity}b.

We next establish a scaling law that estimates the  adsorption length scale $\lambda_i$ without solving Eq.~\eqref{eqn:ad_eigenvalues}. 
Based on dimensional analysis, we propose
\begin{equation}
    \lambdai =
    a\frac{u_0 L^2}{\Da} + b\frac{u_0 L}{\ga}
    \;,
    \label{eq:ad_length_scale}
\end{equation}
where the non-dimensional parameters $a$ and $b$ quantify the influence of diffusivity and adsorption, respectively.
We first estimate $a \approx 5 \cdot 10^{-4}$ from numerical simulations with strong adsorption ($\ga\rightarrow\infty$, implying $\ca=0$ at the walls) and then use simulations with finite adsorption to determine $b\approx 0.03$.
This results in typical length scales of $\unit[5-50]{cm}$ for the parameters given in Table~\ref{table1}.
Taken together, we thus have an analytical prediction of the adsorption length scale~$\lambda_i$, which matches the numerical data for various parameters; see Fig.~\ref{fig2:odorincavity}b.

\begin{figure}
    \centering
    \includegraphics[scale=0.16]{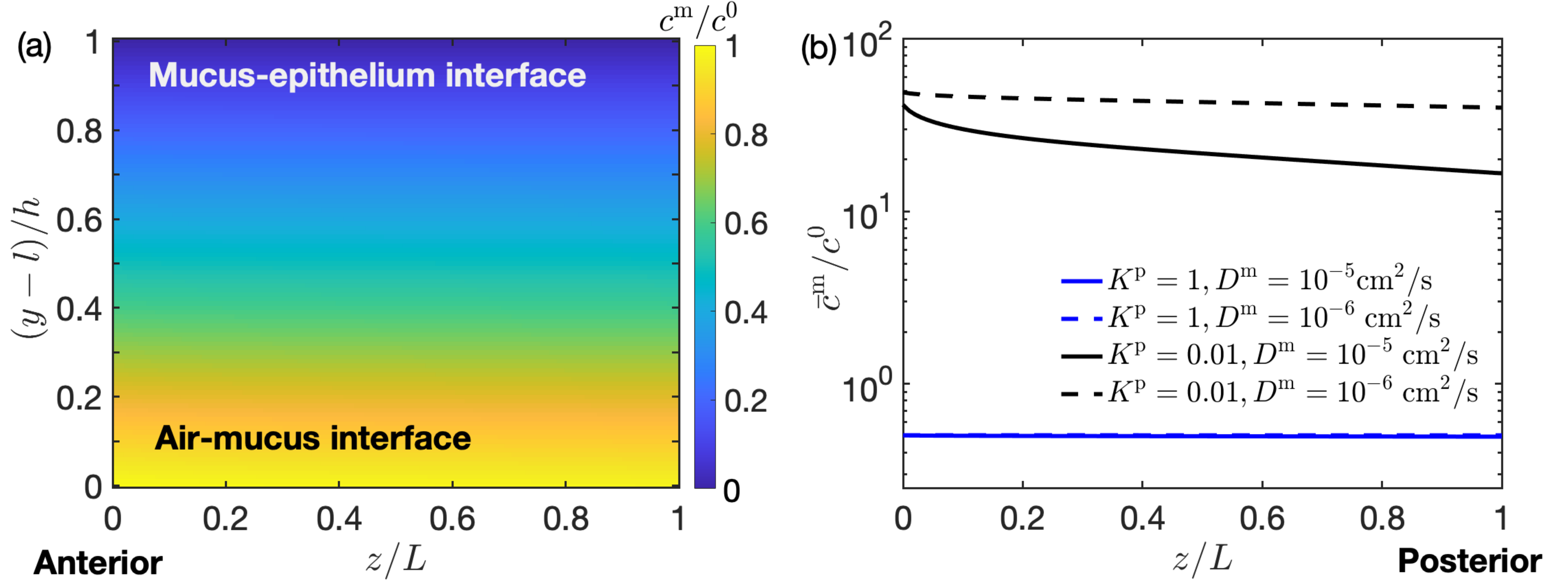}
    \caption{\textbf{Odorant concentrations decrease nearly exponentially inside the mucus.}
    (a) Odorant concentration $\ccm$ in mucus as a function of position~$z$ along the nasal cavity and depth~$y$ in the mucus.
    Data has been obtained from numerical simulations of Eq.~\eqref{eqn:mucus_field} for $\Kpp=1$, $\Dmu=\unitfrac[10^{-5}]{cm^2}{sec}$.
    (b) Cross-sectionally averaged concentration $\cmb$ inside mucus as a function of $z$ for various $\Kpp$ and $\Dmu$.
    (a, b) Model parameters are $\D=\unitfrac[0.15]{cm^2}{s}$, $\g=\unitfrac[10]{cm}{sec}$, $Pe=1000$, $l=\unit[0.3]{cm}$, $h=\unit[0.001]{cm}$, and $L=\unit[10]{cm}$.
    }
    \label{fig3:odorinmucus}
\end{figure}
\label{B}
\subsection{Odorants are absorbed by the aqueous mucus}
\label{sec:mucus}

Odorants adsorbed at the walls of the nasal cavity are actually dissolved in a layer of mucus, which is about $\h \sim \unit[10]{\upmu m }$ thick~\cite{Craven2012,Quraishi1998}.
To estimate the odorant concentration at the receptors located on cilia embedded in the mucus, we next determine the passive transport of odorants in the mucus.
Since mucus only moves slowly ($\unitfrac[1.2-30]{mm}{min}$~\cite{Beule2010}), advection is negligible and the dynamics are dominated by diffusion with diffusivity~$\Dm$.
We thus determine the concentration~$\cm(y, z)$ of the $i$-th odorant in the mucus by solving the stationary diffusion equation.
Odorants generally enter the mucus at the air-mucus interface.
The concentration $\ca(\pm l, z)$ in air at the interface is reduced by a constant factor~$\Ka$ compared to the cross-sectionally averaged concentration~$\cab$ that we estimated in the previous section.
We find $\Ka \approx 2 \Da/(\ga l)$, where the scaling stems from dimensional analysis of the boundary condition given by \Eqref{eqn:advection_diffusion_bc} and we determined the numerical pre-factor from numerical simulations.
The concentration $\cm(\pm l,z)$ dissolved in mucus, $\cm(\pm l,z)=\ca(\pm l,z)/\Kp$, follows from assuming fast dissolution so the concentrations are in local equilibrium.
Here, $\Kp$ determines the odorant solubility in mucus~\cite{Craven2012}, which also includes the effect of odorant-binding proteins~\cite{Jennifer2018, Brito2016, Larter2016}.
On the epithelial side of the mucus, odorants are adsorbed by the highly vascularized bed of epithelial tissue~\cite{strotmann2011,Quraishi1998}.
We again describe this by an adsorption boundary condition, $\pm \Dm \partial_y \cm = \gm\cm$ at $y=\mp(l+\h)$, where $\gm >0$ quantifies the adsorption strength. 
Since the cavity is much longer than the mucus is thick, $L\gg\h$, axial diffusion is negligible compared to the cross-sectional one, implying that the stationary state is governed by $\partial_y^2 \cm =0$.
The analytical solution reads
\begin{equation}
     \cm(y,z)=\frac{\cab(z)}{\Kp\Ka} \left[ 
     1+\frac{(l-y)\gm}{\Dm+ \gm \h}
     \right]
     \;,
     \label{eqn:mucus_field}
 \end{equation}
where  we used $\gm =\ga \Kp$, so the boundary conditions are consistent with those of section \ref{B}.
Fig.~\ref{fig3:odorinmucus}a shows a typical profile inside the mucus.

The odorant receptors are located on cilia that span the entire thickness~$\h$ of the mucus~\cite{DavidPNAS2018,Craven2012}.
We thus estimate the concentration at the receptors using a cross-sectional average, $\bcm(z)=\h^{-1}\int_l^{l+\h} \cm(y,z) \, \diff y$, yielding
\begin{align}
    \bcm(z) &= \cin \alphai e^{-\frac{z}{\lambdai}},
    \label{eqn:mucus_conc}
\end{align}
with    the pre-factor
\begin{align}
    \alphai=\frac{\gm l}{4 \Da (\Kp)^2}\left[2 - \frac{1}{1 + \frac{\Dm}{\gm \h}}\right]
    \;.
    \label{eqn:mucus_conc_prefactor}
\end{align}
Taken together, we showed that the odorant concentration at the olfactory receptors is proportional to the ambient concentration~$c^0_i$ and decays exponentially along the nasal cavity; see Fig.~\ref{fig3:odorinmucus}b.
Note that the square bracket in \Eqref{eqn:mucus_conc_prefactor} is limited to values between $1$ and $2$, implying that the parameters $\Dm$ and $\h$ have limited influence on the odorant concentration~$\bcm$.
The parameters $\gm$, $l$, and $\Da$ can affect the concentration significantly, but the odorant solubility~$\Kp$ and the decay length~$\lambdai$, given by \Eqref{eq:ad_length_scale}, have the strongest influence.

\subsection{Anterior receptors are excited more strongly}
\label{sec:excitation}
We next ask how the distribution of odorant receptors of a particular type affects the total signal that is accumulated in the corresponding glomerulus.
Since we showed that the odorant concentration~$\bcm$ mainly varies along the axial $z$-direction, it suffices to characterize the number density~$n_j(z)$ of receptors of type~$j$ along the same direction.
Each receptor senses many different odorants~\cite{ZwickerPNAS2016,MAINLAND2014,Touhara2009} and in the simplest case the resulting excitation of a receptor of type~$j$ is a linear function of the concentrations of all $\No$ odorants~\cite{ZwickerPNAS2016,TiberiueLife2019}.
We summarize this behavior by a matrix~$S_{ji}$, which quantifies the sensitivity of receptor type~$j$ to odorant~$i$.
Taken together, the excitation of a receptor at position~$z$ is then given by $S_{ji} \bcm(z)$.
The excitation of all receptors of the same type are accumulated in the associated glomerulus, whose excitation~$\Ej$ then reads
\begin{equation}
    \label{eqn:excitation}
    \Ej = \sum_{i=1}^{\No}\Aji \cin + \xi_j
    \;,
\end{equation}
where $\xi_j$ is a random variable describing intrinsic noise and the amplification matrix
\begin{equation}
    \Aji=\Sji \alpha_i \int_0^L n_j(z) e^{-z/\lambda_i} \, \diff z 
    \label{eqn:amplification_factor}
\end{equation}
summarizes the physical processes in the cavity as well as the sensitivity of the receptors.
This simple model approximates the olfactory system as a linear communication channel; see Fig.~\ref{fig1:generalpic}.
The exponential distribution of odorants in the mucus then implies that anteriorly placed receptors generally receive more odorants and exhibit a larger excitation.

\subsection{Transmitted information increases with signal-to-noise ratio of receptors}

The main role of the olfactory system is to measure the chemical composition of the surrounding, i.e., the odor vector $\vect{c} = \{c^0_1, c^0_2, \ldots, c^0_\No\}$, where $c^0_i$ denotes the ambient concentration of odorant~$i$.
To do this, the brain has access to the excitation of  $\Nr \approx 300$ receptor types~\cite{DavidPNAS2018}, summarized in the excitation vector $\vect{e} = \{e_1, e_2, \ldots, e_\Nr\}$.
Both the input $\vect{c}$ and the encoding $\vect{e}$ are high-dimensional vectors, but since $\Nr \ll \No$, the encoding necessarily compresses the information about the odor.
In the simplest case, an optimal encoding preserves as much information about the input as possible, which amounts to maximizing the mutual information~\cite{ITbook1991}
\begin{equation}
   I = \iint \diff\ci \:\diff\boldsymbol{e}\: P(\boldsymbol{\ci}, \boldsymbol{e}) \:\mathrm{log}_2 \left[ \frac {P(\boldsymbol{\ci}, \boldsymbol{e})}{P(\boldsymbol{e})\Penv(\boldsymbol{\ci})}\right]
   \;,
   \label{eqn:mutual_information}
\end{equation}
where $\Penv(\vect{c})$ denotes the probability density of finding odor~$\vect{c}$ in the environment,
$P(\vect{c}, \vect{e})$ is the joint probability density resulting from the mapping given by Eq.~\eqref{eqn:excitation}, and $P(\vect{e})=\int P(\vect{c}, \vect{e}) \, \diff \vect{c}$ is a marginal probability density.
Maximizing the information~$I$ allows studying optimal receptor sensitivities and expression levels~\cite{ZwickerPNAS2016,zwickerPlosone2016,zwickerPloscompbio2019,TiberiueLife2019}.

We here focus on the spatial distribution of the receptors in the olfactory epithelium and ask which configuration leads to the largest information transferal.
For simplicity, we consider normally distributed inputs, $\Penv(\vect{c}) = \norm(\vect\mu, \tens{\Sigma}^\mathrm{c})$, where $\vect\mu$ describes the mean concentration of each odorant and $\vect{\Sigma}^\mathrm{c}$ denotes the covariance matrix.
Moreover, we consider normally distributed intrinsic noise, $P_\xi(\vect\xi) = \norm(\vect0, \tens\Sigma^\xi)$, with zero mean and covariance $\tens\Sigma^\xi$.
The linear mapping given by Eq.~\eqref{eqn:excitation} then implies that the joint distribution $P(\vect{c}, \vect{e})$ is also a normal distribution and the mutual information can be calculated explicitly; see Appendix~\ref{sec:Appendix1}.

To analyze the mutual information~$I$, we for simplicity consider a situation where odor concentrations are uncorrelated, receptors respond to exactly one odorant, and the intrinsic noise is uncorrelated.
These conditions imply that the matrices $\tens\Sigma^\mathrm{c}$, $A_{ji}$, and $\tens\Sigma^\xi$ are diagonal, $I$ becomes independent of $\vect\mu$, and Eq. \eqref{eqn:mutual_information} can be expressed as
\begin{equation}
    I_\mathrm{diag}
    =\frac{1}{2}\sum_{i=1}^\Nr \log_2\Biggl(
        1+\zeta_i^2 \!\left[\int_0^L \!\! n_i(z) e^{-\frac{z}{\lambda_i}} \, \diff z\right]^2
    \Biggr)
    \;,
    \label{eqn:information_uncorr_zeta}
\end{equation}
where $\zeta_i=\alpha_{i} S_{ii}[\SigmaC_{ii}/\Sigma^\xi_{ii}]^{1/2}$ quantifies the signal-to-noise ratio of a single receptor responding to odorant~$i$.
Because odorants are measured independently, the total information~$I$ is the sum of the partial information on each odorant~$i$.
This partial information is a monotonously increasing function~\cite{ITbook1991,zwickerPlosone2016,zwickerPloscompbio2019,TiberiueLife2019} of the signal-to-noise ratio $\zeta_i$. %
Consequently, the transmitted information~$I$ decreases for larger intrinsic noise $\Sigma^\xi_{ii}$.
Conversely, $I$ increases for larger signal (larger variation in the input, quantified by $\SigmaC_{ii}$) and larger intrinsic amplification~$\aii$ of the olfactory system.

We next ask how one could optimize the olfactory system for a given environment.
Since the external variation~$\SigmaC_{ii}$ is determined by the environment, one could either reduce intrinsic noise~$\Sigma^\xi$ or increase amplification factors~$A_{ii}$.
The first option can be achieved by using more or better receptors, but there are biophysical limitations.
The second alternative of improving the amplification factor~$A_{ii}$ is more interesting.
Eq.~\eqref{eqn:amplification_factor} shows that $A_{ii}$ depends on physical parameters (summarized by $\alpha_i$ and $\lambda_i$), the sensitivities $S_{ji}$, and the spatial distribution $n_i(z)$ of receptors.
While physical parameters cannot be easily changed, one could increase the sensitivities~$S_{ji}$ or the total number of receptors to improve~$I$.
However, these two alternatives are analogous to decreasing the intrinsic noise discussed above.
Instead, there is also the option to redistribute the receptors while keeping their total density constant.

\begin{figure}[t]
    \includegraphics[scale=0.46]{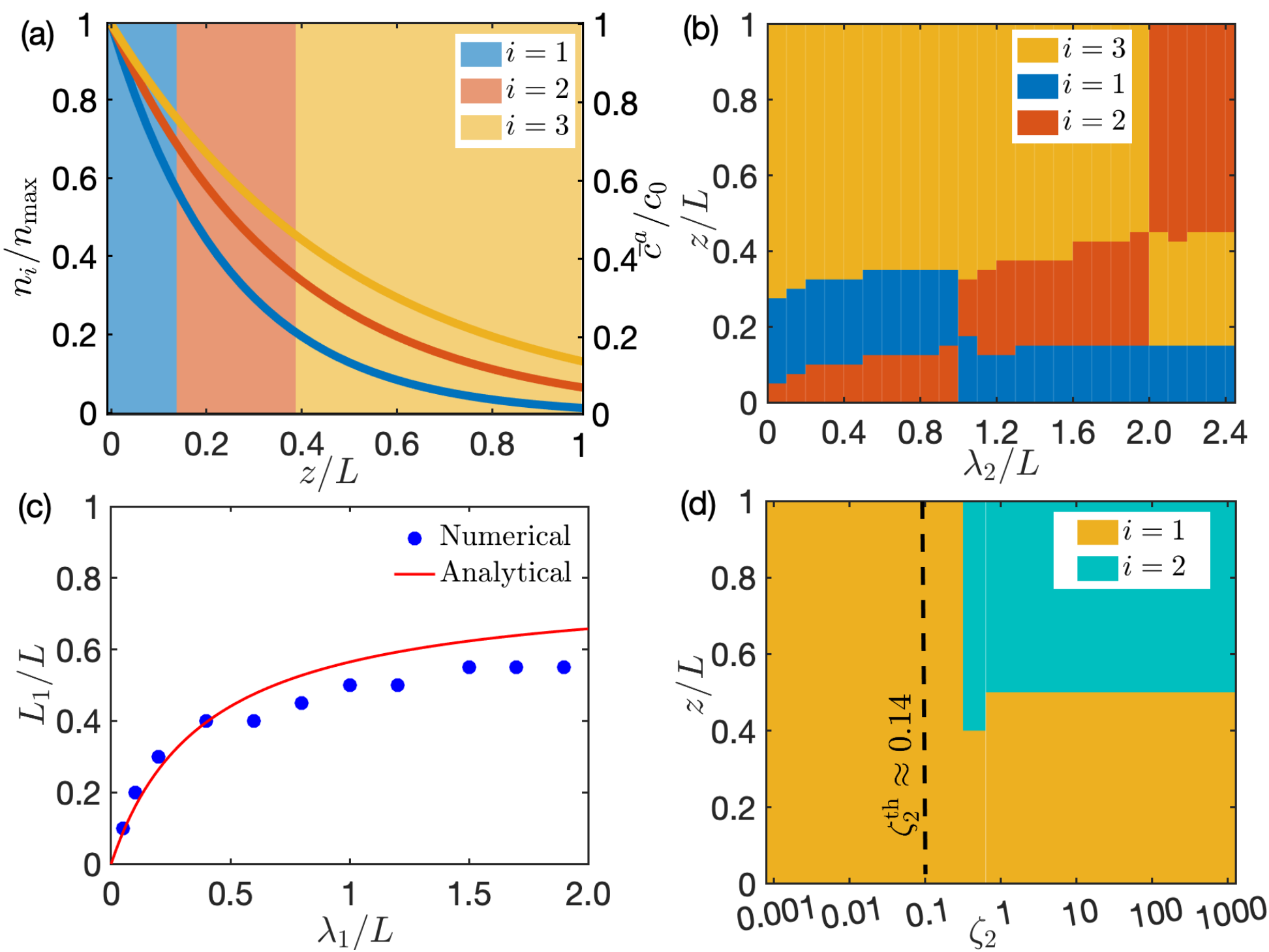}
    \caption{\textbf{Anterior receptors should detect strongly adsorbing odorants.}
    (a) Receptor density $n_i$ (colored regions) as a function of position~$z$ along the cavity obtained from numerical optimization.
    Lines indicate the associated cross-sectionally averaged odorant concentration $\bar c^\mathrm{a}/c_0$ in cavity.
    Receptors are separated into bands, ordered by the absorption length scale of their respective odorants ($\lambda_1=\unit[10]{cm}$, $\lambda_2=\unit[20]{cm}$, and $\lambda_3 = \unit[30]{cm}$).
    (a, b) Model parameters are $L=\unit[10]{cm}$, $\Nr=\No=3$, and we consider the limit of large signal-to-noise where $I$ is given by Eq.~\ref{eqn:high_signal}.
    (b) Receptors zones as a function of the length scale~$\lambda_2$ associated with the second receptor type for $\lambda_1=\unit[10]{cm}$ and $\lambda_3=\unit[20]{cm}$.
    (c) Size~$L_1$ of receptor zone as a function of the length scale~$\lambda_1$ of its odorant for $\lambda_2=\unit[20]{cm}$.
    The analytical theory given by Eq.~\eqref{eqn:information_high_SNR_optimal} (solid lines) explains the data from the numerical optimization (dots).
    (d) Receptor zones as a function of the signal-to-noise ratio~$\zeta_\mathrm{2}$ for $\zeta_1=10$, $\lambda_1=\unit[20]{cm}$, and $\lambda_2=\unit[30]{cm}$.
    The dashed line indicates the estimate for the threshold value  $\zeta_\mathrm{2}^\mathrm{th}$ given by Eq.~\eqref{eqn:zeta_th}.
    (c, d) Model parameters are $L=\unit[10]{cm}$ and $\Nr=\No=2$.
    }
    \label{fig4:receptordist}
\end{figure}

\subsection{Optimal receptor arrays detect strongly adsorbing odorants with anterior receptors}
\label{sec:zonal_distribution}
We finally are in a position to ask which spatial distribution of the odorant receptors in the nasal cavity is optimal.
Since the concentration of odorants absorbed in mucus is largest at the anterior side of the nasal cavity (see Fig.~\ref{fig3:odorinmucus}), it would be optimal to place all receptors there.
However, receptors cannot be packed arbitrarily dense and they would also compete for the same odorant molecules in this case.
Instead, we assume that receptors have a maximal density $\nMax$ and that merely the distributions $n_i(z)$ of individual receptor types can change along the anterior-posterior axis.
For simplicity, we consider the case where the overall density of all receptors reaches $\nMax$ everywhere, although realistic systems likely possess a gradient~\cite{Rosemary2015}.
To see what distributions are optimal, we numerically maximize $I_\mathrm{diag}$ given by Eq.~\eqref{eqn:information_uncorr_zeta}.
Fig.~\ref{fig4:receptordist}a shows that optimal distributions exhibit rectangular functions, where a single receptor type dominates the distribution at each position~$z$ along the nasal cavity.
This optimization thus naturally leads to a zonal distribution of odorant receptors, similar to experimental observations~\cite{Rosemary2015}.
We find that the order and sizes of the receptor zones are reproducible and depend on the physical parameters~$\alpha_i$ and $\lambda_i$.

The physical properties of odorants mainly affect their absorption length scale~$\lambda_i$; see Eq.~\eqref{eq:ad_length_scale}.
Our numerical simulations suggest that strongly absorbing odors (small $\lambda_i$) are detected by receptors at the anterior end of the nasal cavity (small $z$), while receptors for the remaining odorants are posteriorly positioned.
To test this hypothesis, we considered a system of $\No=\Nr=3$ odorant-receptor pairs and systematically varied the absorption length scale of one of the odorants; see Fig.~\ref{fig4:receptordist}b.
These data confirm that the optimal order of the receptors is directly given by the order of the increasing associated absorption length scales.
The data also suggests that the actual lengths of the zones vary only weakly with the absorption lengths.
To understand this behavior in more detail, we investigate the limit of large signal-to-noise ratio ($\zeta_i \gg 1$), where the optimal zone width can be approximated analytically; see Appendix~\ref{sec:Appendix1}.
The resulting expression~\eqref{eqn:information_high_SNR_optimal} agrees with numerical simulations (see Fig.~\ref{fig4:receptordist}c) and suggests that the zone length increases proportionally to the absorption length~$\lambda_i$ for small $\lambda_i$ until it becomes comparable to the lengths of the other zones.
This suggests that there is little advantage in placing receptors at positions beyond the absorption length and it also captures the competition of receptors for space, resulting in an even distribution for similar physical properties.

The physical properties of the receptors mainly affect the sensitivity $S_{ji}$.
Fig.~\ref{fig4:receptordist}d shows that the sensitivity, quantified by the signal-to-noise ratio~$\zeta_i$ of a single receptor, determines whether receptors are useful or not, while the zone length seems to be almost independent of this parameter.
To understand why there is a minimal signal-to-noise ratio~$\zeta_i$ beyond which receptors become useful, we next investigate an array of $\Nr$ similar receptors and ask what signal-to-noise ratio~$\zeta_\mathrm{add}$ an additional receptor needs to have to become useful, i.e., to increase the overall information~$I$.
For simplicity, we consider large absorption lengths ($\lambda_i \rightarrow \infty$) where only the total count~$N_i$ of receptors of type~$i$ matters while their spatial distribution is insignificant.
Eq.~\eqref{eqn:zeta_th} in the appendix gives the analytical value for the threshold value, which approximates the numerically determined one; see Fig.~\ref{fig4:receptordist}d.
The analytical expression demonstrates that additional receptor types can only be beneficial if $N\zeta > \Nr$.
Since $N/\Nr$ is the mean number of receptors per type, this bound indicates that the average signal-to-noise of the accumulated response for each type needs to exceed~$1$ before adding a new type can be advantageous.
Beyond this threshold, additional receptor types quickly contribute to the overall information, even if their signal-to-noise ratio~$\zeta_\mathrm{add}$ is less than the average value of receptor types already present in the system.
Taken together, this analysis provides qualitative guidance for a relation between receptor sensitivity~$\zeta$, receptor repertoire size~$\Nr$, and the total count~$N$ of receptors in the nasal cavity.

\section{Discussion}

We proposed a comprehensive model that describes how odorant molecules distribute in the nasal cavity and excite olfactory receptors.
In particular, we developed a physical theory that predicts the distribution of odorants in the mucus layer of the nasal cavity, taking into account physical transport and adsorption processes.
We found that the concentration of odorants decays exponentially with the distance from the anterior side and that the overall magnitude of the dissolved odorants is mainly governed by their solubility.
Using information theory, we then concluded that this odorant distribution informs how receptors should be distributed in the olfactory epithelium for maximal information transferal to the brain.
In particular, we found that olfactory receptors segregate into different zones, similar to experimental observations~\cite{Horowitz12241,tanchemsense2018,AxelCell1993}, with receptors sensitive to more soluble odorants located anteriorly.
Taken together, our model suggest that the physical transport of odorants in the nasal cavity affects their distribution in mucus significantly.
Olfactory receptors and the downstream processing need to take this into account to be able to reconstruct a faithful representation of the ambient odor.

To obtain insight and analytical results, our model employed several simplifications.
While the qualitative results will likely prevail, the quantitative details might change when some of the simplifications are lifted.
For instance, odorant transport in realistic nasal geometries might be altered and affect receptor placement, which could be analyzed using detailed computational fluid dynamics modeling~\cite{Barbarite2021, bruning2020, Chengyu2018}.
Moreover, the linear encoding that we study is a simple abstraction of the odorant detection by receptors since real receptors have non-linear dose-response curves~\cite{Gabriel2020} and the downstream processing of the excitation is also non-linear~\cite{zwickerPlosone2016,zwickerPloscompbio2019,BalasubramanianPNAS2019, Shanshan2019, Reddy2018}.
Incorporating non-linearities in the theory will likely change details, like the width of receptor zones, but receptors responding to more soluble odors still need to be placed anteriorly.
Receptors also respond to many different odorants, so their placement should be based on the concentration profiles averaged across all odorants they respond to.
Here, receptors that predominately respond to odorants originating from food might be preferentially placed posteriorly to aid in retronasal smell.
We hypothesize that the competition between orthonasal and retronasal smell, the fact that receptors detect many odorants, and the physical limitations of the genetic circuit that selects receptor identity of olfactory neurons~\cite{colemanJNeu2019, horgue2022} lead to much less well defined receptor zones than our theory predicts.
However, a heterogeneous distribution of receptors can likely increase overall information transmission, which might also aid in improving artificial olfactory devices~\cite{Raman2011}.

\begin{acknowledgments}
We thank Lucas Menou,  Ajinkya Kulkarni, and Jan Kirschbaum for helpful discussions and we gratefully acknowledge funding from the Max Planck Society.%
\end{acknowledgments}

\appendix

\section{Mutual information}
\label{sec:mutual_information}
We here analyze the mutual information~$I$, given by \Eqref{eqn:mutual_information}, for uncorrelated inputs and noises. %
Assuming Gaussian input and noise distributions, $\Penv(\vect{c}) = \norm(\vect\mu, \tens{\Sigma}^\mathrm{c})$ and $P_\xi(\vect\xi) = \norm(\vect0, \tens\Sigma^\xi)$, the distribution of the output $\vect{e}$, defined by the linear map given in \Eqref{eqn:excitation}, is also Gaussian and reads $P(\vect{e}) = \norm(\tens{A} \vect\mu^\mathrm{c}, \tens\Sigma^\mathrm{e} )$ with $\tens\Sigma^\mathrm{e}=\tens{A}\tens{\Sigma}^\mathrm{c}\tens{A}^{T}+\tens{\Sigma}^\mathrm{\xi}$~\cite{tong1990multivariate}.
Similarly, we find a Gaussian joint distribution $P(\ci,\vect{e})=\norm(\vect\mu,\tens\Sigma)$ with
\begin{align}
	\vect\mu&=\begin{bmatrix} 
	\vect\mu^\mathrm{c} \\
	\tens A \vect \mu^\mathrm{c}
	\end{bmatrix}
&\text{and} &&
    \tens\Sigma
   =\begin{bmatrix}
    \tens\Sigma^\mathrm{c} &
    \tens\Sigma^\mathrm{c}\tens{A}^{T} \\ 
    \tens{A}\tens\Sigma^\mathrm{c} & \tens{A}\tens\Sigma^\mathrm{c}\tens{A}^{T}+\tens\Sigma^{\xi}
    \end{bmatrix}
    \;.
\end{align}
Hence, the mutual information reduces to~\cite{Forney1998}
\begin{equation}
    I=\frac{1}{2}\log_2\left(\frac{|\tens\Sigma^\mathrm{c}||\tens\Sigma^\mathrm{e}|}{|\tens\Sigma|}\right)
    \;.
\end{equation}
For the simple case where the matrices $\tens\Sigma^\mathrm{c}$, $\tens A$, and $\tens\Sigma^\xi$ are diagonal, we have
\begin{equation}
    I_\mathrm{diag}
    =\frac{1}{2}\sum_{i=1}^\Nr \log_2\left(1+\aii^2 \frac{\SigmaC_{ii}}{\Sigma^\xi_{ii}} \right)
    \;.
    \label{eqn:information_uncorr}
\end{equation}
Using the definition of $A_{ii}$ given in Eq.~\eqref{eqn:amplification_factor}, we find Eq.~\eqref{eqn:information_uncorr_zeta}
where $\zeta_i=\alpha_{i} S_{ii}[\SigmaC_{ii}/\Sigma^\xi_{ii}]^{1/2}$ quantifies the signal-to-noise ratio of a single receptor.

We start by considering the case of high signal-to-noise ($\zeta_i \gg 1$). In this case, Eq. \eqref{eqn:information_uncorr_zeta} becomes
\begin{align}
   I_{\mathrm{diag, H}} \approx
   \sum_{i=1}^{\Nr} \log_2  \zeta_i +
   \sum_{i=1}^{\Nr} \log_2 \left(\int_0^L \!\! n_i e^{-\frac{z}{\lambda_i}} \, \diff z\right)
    \label{eqn:high_signal}
    \;,
\end{align}
so we can maximize the information irrespective of $\zeta_i$.
In particular, we consider the situation where receptors are sorted into distinct zones and optimize~$I_{\mathrm{diag, H}}$ with respect to the zone length.
In the simplest case of $\Nr=2$ types of receptors, we have
\begin{subequations}
\begin{align}
    n_1(z) &= \begin{cases}\nMax &  0<z<L_1 \\ 0 & \text{otherwise}\end{cases}
\\ 
    n_2(z) &= \begin{cases}\nMax &  L_1<z<L \\ 0 & \text{otherwise}\end{cases}
\end{align}
\end{subequations}
where we assumed $\lambda_1 < \lambda_2$ and $\nMax=N/L$.
Optimizing $I_{\mathrm{diag, H}}$ with respect to $L_1$, we find the optimal lengths
\begin{equation}
    L_{1}^*(\lambda_1,\lambda_2)
    =\frac{2\lambda_1\lambda_2(1-e^{L/\lambda_2})}{\lambda_2-(2\lambda_1+\lambda_2)e^{L/\lambda_2}}
    \approx \frac{2\lambda_1\lambda_2}{2\lambda_1+\lambda_2}
     \label{eqn:information_high_SNR_optimal} 
\end{equation}
and $L^*_2 = L - L^*_1$.

In the opposite limit of low signal-to-noise ($\zeta_i N_i \ll 1$) we consider for simplicity  $\lambda_i\rightarrow\infty$, so Eq. \eqref{eqn:information_uncorr_zeta} becomes
\begin{equation}
    \label{eqn:Nrtype}
    I_\mathrm{diag,L} \approx \frac{1}{2\ln 2} \sum_i N_i^2\zeta_i^2
    \;,
\end{equation}
where the total number of receptors is $N=\sum_iN_i$.
We then ask when it is useful to add an additional receptor type to an existing array.
For simplicity, we consider an array of $\Nr$ receptor types with equal $\zeta_i=\zeta$ for $i=1,\ldots,\Nr$ and add a receptor type with potentially different signal-to-noise $\zeta_\mathrm{add}$.
Using $N_i = (N - N_\mathrm{add})/\Nr$ for $i=1,\ldots,\Nr$, we obtain the optimal count of the added receptor, $N^*_\mathrm{add}(N, \zeta, \zeta_\mathrm{add})$, from the optimality condition $\partial I_\mathrm{diag, L} / \partial N_\mathrm{add} = 0$,
\begin{align}
   \label{eqn:Nadd}
N^*_\mathrm{add} &= 
    \frac{\Nr N \zeta^2(\Nr^2+N^2\zeta^2)}
    {\Nr \zeta^2 (\Nr^2 -N^2 \zeta^2) +\zeta_\mathrm{add}^2(\Nr^2 +N^2 \zeta^2)^2}
    \;,
\end{align}
which is valid for $\zeta_\mathrm{add} > \zeta^{\mathrm{th}}_\mathrm{add}$.
The threshold value~$\zeta^{\mathrm{th}}_\mathrm{add}$ follows from the pole of \Eqref{eqn:Nadd} and reads
\begin{equation}
    \zeta^{\mathrm{th}}_\mathrm{add}
    =\frac{\zeta\sqrt{N^2\Nr \zeta^2 - \Nr^3}}
    {N^2\zeta^2 + \Nr^2}
   \;.
   \label{eqn:zeta_th}
\end{equation}
Since $\zeta^{\mathrm{th}}_\mathrm{add}$ needs to be real, we require $N\zeta > \Nr$.

\label{sec:Appendix1}

\bibliography{references}

\begin{thebibliography}{43}%
\makeatletter
\providecommand \@ifxundefined [1]{%
 \@ifx{#1\undefined}
}%
\providecommand \@ifnum [1]{%
 \ifnum #1\expandafter \@firstoftwo
 \else \expandafter \@secondoftwo
 \fi
}%
\providecommand \@ifx [1]{%
 \ifx #1\expandafter \@firstoftwo
 \else \expandafter \@secondoftwo
 \fi
}%
\providecommand \natexlab [1]{#1}%
\providecommand \enquote  [1]{``#1''}%
\providecommand \bibnamefont  [1]{#1}%
\providecommand \bibfnamefont [1]{#1}%
\providecommand \citenamefont [1]{#1}%
\providecommand \href@noop [0]{\@secondoftwo}%
\providecommand \href [0]{\begingroup \@sanitize@url \@href}%
\providecommand \@href[1]{\@@startlink{#1}\@@href}%
\providecommand \@@href[1]{\endgroup#1\@@endlink}%
\providecommand \@sanitize@url [0]{\catcode `\\12\catcode `\$12\catcode
  `\&12\catcode `\#12\catcode `\^12\catcode `\_12\catcode `\%12\relax}%
\providecommand \@@startlink[1]{}%
\providecommand \@@endlink[0]{}%
\providecommand \url  [0]{\begingroup\@sanitize@url \@url }%
\providecommand \@url [1]{\endgroup\@href {#1}{\urlprefix }}%
\providecommand \urlprefix  [0]{URL }%
\providecommand \Eprint [0]{\href }%
\providecommand \doibase [0]{http://dx.doi.org/}%
\providecommand \selectlanguage [0]{\@gobble}%
\providecommand \bibinfo  [0]{\@secondoftwo}%
\providecommand \bibfield  [0]{\@secondoftwo}%
\providecommand \translation [1]{[#1]}%
\providecommand \BibitemOpen [0]{}%
\providecommand \bibitemStop [0]{}%
\providecommand \bibitemNoStop [0]{.\EOS\space}%
\providecommand \EOS [0]{\spacefactor3000\relax}%
\providecommand \BibitemShut  [1]{\csname bibitem#1\endcsname}%
\let\auto@bib@innerbib\@empty
%</preamble>
\bibitem [{\citenamefont {Silva~Teixeira}\ \emph {et~al.}(2016)\citenamefont
  {Silva~Teixeira}, \citenamefont {Cerqueira},\ and\ \citenamefont
  {Silva~Ferreira}}]{silvachemsense2016}%
  \BibitemOpen
  \bibfield  {author} {\bibinfo {author} {\bibfnamefont {C.~S.}\ \bibnamefont
  {Silva~Teixeira}}, \bibinfo {author} {\bibfnamefont {N.~M.}\ \bibnamefont
  {Cerqueira}}, \ and\ \bibinfo {author} {\bibfnamefont {A.~C.}\ \bibnamefont
  {Silva~Ferreira}},\ }\href@noop {} {\bibfield  {journal} {\bibinfo  {journal}
  {Chemical Senses}\ }\textbf {\bibinfo {volume} {41}},\ \bibinfo {pages} {105}
  (\bibinfo {year} {2016})}\BibitemShut {NoStop}%
\bibitem [{\citenamefont {Knudsen}\ \emph {et~al.}(1993)\citenamefont
  {Knudsen}, \citenamefont {Tollsten},\ and\ \citenamefont
  {Bergström}}]{KnudsenPhytochemistry1993}%
  \BibitemOpen
  \bibfield  {author} {\bibinfo {author} {\bibfnamefont {J.~T.}\ \bibnamefont
  {Knudsen}}, \bibinfo {author} {\bibfnamefont {L.}~\bibnamefont {Tollsten}}, \
  and\ \bibinfo {author} {\bibfnamefont {L.}~\bibnamefont {Bergström}},\
  }\href {\doibase https://doi.org/10.1016/0031-9422(93)85502-I} {\bibfield
  {journal} {\bibinfo  {journal} {Phytochemistry}\ }\textbf {\bibinfo {volume}
  {33}},\ \bibinfo {pages} {253} (\bibinfo {year} {1993})},\ \bibinfo {note}
  {the International Journal of Plant Biochemistry}\BibitemShut {NoStop}%
\bibitem [{\citenamefont {Mainland}\ \emph {et~al.}(2014)\citenamefont
  {Mainland}, \citenamefont {Lundström}, \citenamefont {Reisert},\ and\
  \citenamefont {Lowe}}]{MAINLAND2014}%
  \BibitemOpen
  \bibfield  {author} {\bibinfo {author} {\bibfnamefont {J.~D.}\ \bibnamefont
  {Mainland}}, \bibinfo {author} {\bibfnamefont {J.~N.}\ \bibnamefont
  {Lundström}}, \bibinfo {author} {\bibfnamefont {J.}~\bibnamefont {Reisert}},
  \ and\ \bibinfo {author} {\bibfnamefont {G.}~\bibnamefont {Lowe}},\ }\href
  {\doibase https://doi.org/10.1016/j.tins.2014.05.005} {\bibfield  {journal}
  {\bibinfo  {journal} {Trends in Neurosciences}\ }\textbf {\bibinfo {volume}
  {37}},\ \bibinfo {pages} {443} (\bibinfo {year} {2014})}\BibitemShut
  {NoStop}%
\bibitem [{\citenamefont {Touhara}\ and\ \citenamefont
  {Vosshall}(2009)}]{Touhara2009}%
  \BibitemOpen
  \bibfield  {author} {\bibinfo {author} {\bibfnamefont {K.}~\bibnamefont
  {Touhara}}\ and\ \bibinfo {author} {\bibfnamefont {L.~B.}\ \bibnamefont
  {Vosshall}},\ }\href {\doibase 10.1146/annurev.physiol.010908.163209}
  {\bibfield  {journal} {\bibinfo  {journal} {Annual Review of Physiology}\
  }\textbf {\bibinfo {volume} {71}},\ \bibinfo {pages} {307} (\bibinfo {year}
  {2009})},\ \bibinfo {note} {pMID: 19575682},\ \Eprint
  {http://arxiv.org/abs/https://doi.org/10.1146/annurev.physiol.010908.163209}
  {https://doi.org/10.1146/annurev.physiol.010908.163209} \BibitemShut
  {NoStop}%
\bibitem [{\citenamefont {Atick}(2011)}]{Atick2011}%
  \BibitemOpen
  \bibfield  {author} {\bibinfo {author} {\bibfnamefont {J.~J.}\ \bibnamefont
  {Atick}},\ }\href {\doibase 10.3109/0954898X.2011.638888} {\bibfield
  {journal} {\bibinfo  {journal} {Network: Computation in Neural Systems}\
  }\textbf {\bibinfo {volume} {22}},\ \bibinfo {pages} {4} (\bibinfo {year}
  {2011})},\ \bibinfo {note} {pMID: 22149669},\ \Eprint
  {http://arxiv.org/abs/https://doi.org/10.3109/0954898X.2011.638888}
  {https://doi.org/10.3109/0954898X.2011.638888} \BibitemShut {NoStop}%
\bibitem [{\citenamefont {Tka\v{c}ik}\ and\ \citenamefont
  {Bialek}(2016)}]{Bialek2016}%
  \BibitemOpen
  \bibfield  {author} {\bibinfo {author} {\bibfnamefont {G.}~\bibnamefont
  {Tka\v{c}ik}}\ and\ \bibinfo {author} {\bibfnamefont {W.}~\bibnamefont
  {Bialek}},\ }\href {\doibase 10.1146/annurev-conmatphys-031214-014803}
  {\bibfield  {journal} {\bibinfo  {journal} {Annual Review of Condensed Matter
  Physics}\ }\textbf {\bibinfo {volume} {7}},\ \bibinfo {pages} {89} (\bibinfo
  {year} {2016})},\ \Eprint
  {http://arxiv.org/abs/https://doi.org/10.1146/annurev-conmatphys-031214-014803}
  {https://doi.org/10.1146/annurev-conmatphys-031214-014803} \BibitemShut
  {NoStop}%
\bibitem [{\citenamefont {Teşileanu}\ \emph {et~al.}(2019)\citenamefont
  {Teşileanu}, \citenamefont {Cocco}, \citenamefont {Monasson},\ and\
  \citenamefont {Balasubramanian}}]{TiberiueLife2019}%
  \BibitemOpen
  \bibfield  {author} {\bibinfo {author} {\bibfnamefont {T.}~\bibnamefont
  {Teşileanu}}, \bibinfo {author} {\bibfnamefont {S.}~\bibnamefont {Cocco}},
  \bibinfo {author} {\bibfnamefont {R.}~\bibnamefont {Monasson}}, \ and\
  \bibinfo {author} {\bibfnamefont {V.}~\bibnamefont {Balasubramanian}},\
  }\href {\doibase 10.7554/eLife.39279} {\bibfield  {journal} {\bibinfo
  {journal} {eLife}\ }\textbf {\bibinfo {volume} {8}},\ \bibinfo {pages}
  {e39279} (\bibinfo {year} {2019})}\BibitemShut {NoStop}%
\bibitem [{\citenamefont {Zwicker}\ \emph {et~al.}(2016)\citenamefont
  {Zwicker}, \citenamefont {Murugan},\ and\ \citenamefont
  {Brenner}}]{ZwickerPNAS2016}%
  \BibitemOpen
  \bibfield  {author} {\bibinfo {author} {\bibfnamefont {D.}~\bibnamefont
  {Zwicker}}, \bibinfo {author} {\bibfnamefont {A.}~\bibnamefont {Murugan}}, \
  and\ \bibinfo {author} {\bibfnamefont {M.~P.}\ \bibnamefont {Brenner}},\
  }\href {\doibase 10.1073/pnas.1600357113} {\bibfield  {journal} {\bibinfo
  {journal} {Proceedings of the National Academy of Sciences}\ }\textbf
  {\bibinfo {volume} {113}},\ \bibinfo {pages} {5570} (\bibinfo {year}
  {2016})},\ \Eprint
  {http://arxiv.org/abs/https://www.pnas.org/content/113/20/5570.full.pdf}
  {https://www.pnas.org/content/113/20/5570.full.pdf} \BibitemShut {NoStop}%
\bibitem [{\citenamefont {Tan}\ and\ \citenamefont
  {Xie}(2018)}]{tanchemsense2018}%
  \BibitemOpen
  \bibfield  {author} {\bibinfo {author} {\bibfnamefont {L.}~\bibnamefont
  {Tan}}\ and\ \bibinfo {author} {\bibfnamefont {X.~S.}\ \bibnamefont {Xie}},\
  }\href@noop {} {\bibfield  {journal} {\bibinfo  {journal} {Chemical senses}\
  }\textbf {\bibinfo {volume} {43}},\ \bibinfo {pages} {427} (\bibinfo {year}
  {2018})}\BibitemShut {NoStop}%
\bibitem [{\citenamefont {Horowitz}\ \emph {et~al.}(2014)\citenamefont
  {Horowitz}, \citenamefont {Saraiva}, \citenamefont {Kuang}, \citenamefont
  {Yoon},\ and\ \citenamefont {Buck}}]{Horowitz12241}%
  \BibitemOpen
  \bibfield  {author} {\bibinfo {author} {\bibfnamefont {L.~F.}\ \bibnamefont
  {Horowitz}}, \bibinfo {author} {\bibfnamefont {L.~R.}\ \bibnamefont
  {Saraiva}}, \bibinfo {author} {\bibfnamefont {D.}~\bibnamefont {Kuang}},
  \bibinfo {author} {\bibfnamefont {K.-h.}\ \bibnamefont {Yoon}}, \ and\
  \bibinfo {author} {\bibfnamefont {L.~B.}\ \bibnamefont {Buck}},\ }\href
  {\doibase 10.1523/JNEUROSCI.1779-14.2014} {\bibfield  {journal} {\bibinfo
  {journal} {Journal of Neuroscience}\ }\textbf {\bibinfo {volume} {34}},\
  \bibinfo {pages} {12241} (\bibinfo {year} {2014})},\ \Eprint
  {http://arxiv.org/abs/https://www.jneurosci.org/content/34/37/12241.full.pdf}
  {https://www.jneurosci.org/content/34/37/12241.full.pdf} \BibitemShut
  {NoStop}%
\bibitem [{\citenamefont {Vedin}\ \emph {et~al.}(2009)\citenamefont {Vedin},
  \citenamefont {Molander}, \citenamefont {Bohm},\ and\ \citenamefont
  {Berghard}}]{vedinJCN2009}%
  \BibitemOpen
  \bibfield  {author} {\bibinfo {author} {\bibfnamefont {V.}~\bibnamefont
  {Vedin}}, \bibinfo {author} {\bibfnamefont {M.}~\bibnamefont {Molander}},
  \bibinfo {author} {\bibfnamefont {S.}~\bibnamefont {Bohm}}, \ and\ \bibinfo
  {author} {\bibfnamefont {A.}~\bibnamefont {Berghard}},\ }\href@noop {}
  {\bibfield  {journal} {\bibinfo  {journal} {Journal of Comparative
  Neurology}\ }\textbf {\bibinfo {volume} {513}},\ \bibinfo {pages} {375}
  (\bibinfo {year} {2009})}\BibitemShut {NoStop}%
\bibitem [{\citenamefont {Vassar}\ \emph {et~al.}(1993)\citenamefont {Vassar},
  \citenamefont {Ngai},\ and\ \citenamefont {Axel}}]{AxelCell1993}%
  \BibitemOpen
  \bibfield  {author} {\bibinfo {author} {\bibfnamefont {R.}~\bibnamefont
  {Vassar}}, \bibinfo {author} {\bibfnamefont {J.}~\bibnamefont {Ngai}}, \ and\
  \bibinfo {author} {\bibfnamefont {R.}~\bibnamefont {Axel}},\ }\href {\doibase
  https://doi.org/10.1016/0092-8674(93)90422-M} {\bibfield  {journal} {\bibinfo
   {journal} {Cell}\ }\textbf {\bibinfo {volume} {74}},\ \bibinfo {pages} {309}
  (\bibinfo {year} {1993})}\BibitemShut {NoStop}%
\bibitem [{\citenamefont {Ressler}\ \emph {et~al.}(1993)\citenamefont
  {Ressler}, \citenamefont {Sullivan},\ and\ \citenamefont
  {Buck}}]{Resslercell1993}%
  \BibitemOpen
  \bibfield  {author} {\bibinfo {author} {\bibfnamefont {K.~J.}\ \bibnamefont
  {Ressler}}, \bibinfo {author} {\bibfnamefont {S.~L.}\ \bibnamefont
  {Sullivan}}, \ and\ \bibinfo {author} {\bibfnamefont {L.~B.}\ \bibnamefont
  {Buck}},\ }\href@noop {} {\bibfield  {journal} {\bibinfo  {journal} {Cell}\
  }\textbf {\bibinfo {volume} {73}},\ \bibinfo {pages} {597} (\bibinfo {year}
  {1993})}\BibitemShut {NoStop}%
\bibitem [{\citenamefont {Barbarite}\ \emph {et~al.}(2021)\citenamefont
  {Barbarite}, \citenamefont {Gadkaree}, \citenamefont {Melchionna},
  \citenamefont {Zwicker},\ and\ \citenamefont {Lindsay}}]{Barbarite2021}%
  \BibitemOpen
  \bibfield  {author} {\bibinfo {author} {\bibfnamefont {E.}~\bibnamefont
  {Barbarite}}, \bibinfo {author} {\bibfnamefont {S.~K.}\ \bibnamefont
  {Gadkaree}}, \bibinfo {author} {\bibfnamefont {S.}~\bibnamefont
  {Melchionna}}, \bibinfo {author} {\bibfnamefont {D.}~\bibnamefont {Zwicker}},
  \ and\ \bibinfo {author} {\bibfnamefont {R.~W.}\ \bibnamefont {Lindsay}},\
  }\href@noop {} {\bibfield  {journal} {\bibinfo  {journal} {Plastic and
  Reconstructive Surgery}\ }\textbf {\bibinfo {volume} {148}},\ \bibinfo
  {pages} {592e} (\bibinfo {year} {2021})}\BibitemShut {NoStop}%
\bibitem [{\citenamefont {Br{\"u}ning}\ \emph {et~al.}(2020)\citenamefont
  {Br{\"u}ning}, \citenamefont {Hildebrandt}, \citenamefont {Heppt},
  \citenamefont {Schmidt}, \citenamefont {Lamecker}, \citenamefont {Szengel},
  \citenamefont {Amiridze}, \citenamefont {Ramm}, \citenamefont {Bindernagel},
  \citenamefont {Zachow} \emph {et~al.}}]{bruning2020}%
  \BibitemOpen
  \bibfield  {author} {\bibinfo {author} {\bibfnamefont {J.}~\bibnamefont
  {Br{\"u}ning}}, \bibinfo {author} {\bibfnamefont {T.}~\bibnamefont
  {Hildebrandt}}, \bibinfo {author} {\bibfnamefont {W.}~\bibnamefont {Heppt}},
  \bibinfo {author} {\bibfnamefont {N.}~\bibnamefont {Schmidt}}, \bibinfo
  {author} {\bibfnamefont {H.}~\bibnamefont {Lamecker}}, \bibinfo {author}
  {\bibfnamefont {A.}~\bibnamefont {Szengel}}, \bibinfo {author} {\bibfnamefont
  {N.}~\bibnamefont {Amiridze}}, \bibinfo {author} {\bibfnamefont
  {H.}~\bibnamefont {Ramm}}, \bibinfo {author} {\bibfnamefont {M.}~\bibnamefont
  {Bindernagel}}, \bibinfo {author} {\bibfnamefont {S.}~\bibnamefont {Zachow}},
   \emph {et~al.},\ }\href@noop {} {\bibfield  {journal} {\bibinfo  {journal}
  {Scientific Reports}\ }\textbf {\bibinfo {volume} {10}},\ \bibinfo {pages}
  {1} (\bibinfo {year} {2020})}\BibitemShut {NoStop}%
\bibitem [{\citenamefont {Li}\ \emph {et~al.}(2018)\citenamefont {Li},
  \citenamefont {Jiang}, \citenamefont {Kim}, \citenamefont {Otto},
  \citenamefont {Farag}, \citenamefont {Cowart}, \citenamefont {Pribitkin},
  \citenamefont {Dalton},\ and\ \citenamefont {Zhao}}]{Chengyu2018}%
  \BibitemOpen
  \bibfield  {author} {\bibinfo {author} {\bibfnamefont {C.}~\bibnamefont
  {Li}}, \bibinfo {author} {\bibfnamefont {J.}~\bibnamefont {Jiang}}, \bibinfo
  {author} {\bibfnamefont {K.}~\bibnamefont {Kim}}, \bibinfo {author}
  {\bibfnamefont {B.~A.}\ \bibnamefont {Otto}}, \bibinfo {author}
  {\bibfnamefont {A.~A.}\ \bibnamefont {Farag}}, \bibinfo {author}
  {\bibfnamefont {B.~J.}\ \bibnamefont {Cowart}}, \bibinfo {author}
  {\bibfnamefont {E.~A.}\ \bibnamefont {Pribitkin}}, \bibinfo {author}
  {\bibfnamefont {P.}~\bibnamefont {Dalton}}, \ and\ \bibinfo {author}
  {\bibfnamefont {K.}~\bibnamefont {Zhao}},\ }\href {\doibase
  10.1093/chemse/bjy013} {\bibfield  {journal} {\bibinfo  {journal} {Chemical
  Senses}\ }\textbf {\bibinfo {volume} {43}},\ \bibinfo {pages} {229} (\bibinfo
  {year} {2018})},\ \Eprint
  {http://arxiv.org/abs/https://academic.oup.com/chemse/article-pdf/43/4/229/25084123/bjy013.pdf}
  {https://academic.oup.com/chemse/article-pdf/43/4/229/25084123/bjy013.pdf}
  \BibitemShut {NoStop}%
\bibitem [{\citenamefont {Zwicker}\ \emph {et~al.}(2018)\citenamefont
  {Zwicker}, \citenamefont {Ostilla-M{\'o}nico}, \citenamefont {Lieberman},\
  and\ \citenamefont {Brenner}}]{DavidPNAS2018}%
  \BibitemOpen
  \bibfield  {author} {\bibinfo {author} {\bibfnamefont {D.}~\bibnamefont
  {Zwicker}}, \bibinfo {author} {\bibfnamefont {R.}~\bibnamefont
  {Ostilla-M{\'o}nico}}, \bibinfo {author} {\bibfnamefont {D.~E.}\ \bibnamefont
  {Lieberman}}, \ and\ \bibinfo {author} {\bibfnamefont {M.~P.}\ \bibnamefont
  {Brenner}},\ }\href {\doibase 10.1073/pnas.1714795115} {\bibfield  {journal}
  {\bibinfo  {journal} {Proceedings of the National Academy of Sciences}\
  }\textbf {\bibinfo {volume} {115}},\ \bibinfo {pages} {2936} (\bibinfo {year}
  {2018})},\ \Eprint
  {http://arxiv.org/abs/https://www.pnas.org/content/115/12/2936.full.pdf}
  {https://www.pnas.org/content/115/12/2936.full.pdf} \BibitemShut {NoStop}%
\bibitem [{\citenamefont {Smear}\ \emph {et~al.}(2011)\citenamefont {Smear},
  \citenamefont {Shusterman}, \citenamefont {O’Connor}, \citenamefont
  {Bozza},\ and\ \citenamefont {Rinberg}}]{smear2011}%
  \BibitemOpen
  \bibfield  {author} {\bibinfo {author} {\bibfnamefont {M.}~\bibnamefont
  {Smear}}, \bibinfo {author} {\bibfnamefont {R.}~\bibnamefont {Shusterman}},
  \bibinfo {author} {\bibfnamefont {R.}~\bibnamefont {O’Connor}}, \bibinfo
  {author} {\bibfnamefont {T.}~\bibnamefont {Bozza}}, \ and\ \bibinfo {author}
  {\bibfnamefont {D.}~\bibnamefont {Rinberg}},\ }\href@noop {} {\bibfield
  {journal} {\bibinfo  {journal} {Nature}\ }\textbf {\bibinfo {volume} {479}},\
  \bibinfo {pages} {397} (\bibinfo {year} {2011})}\BibitemShut {NoStop}%
\bibitem [{\citenamefont {Shusterman}\ \emph {et~al.}(2011)\citenamefont
  {Shusterman}, \citenamefont {Smear}, \citenamefont {Koulakov},\ and\
  \citenamefont {Rinberg}}]{shusterman2011}%
  \BibitemOpen
  \bibfield  {author} {\bibinfo {author} {\bibfnamefont {R.}~\bibnamefont
  {Shusterman}}, \bibinfo {author} {\bibfnamefont {M.~C.}\ \bibnamefont
  {Smear}}, \bibinfo {author} {\bibfnamefont {A.~A.}\ \bibnamefont {Koulakov}},
  \ and\ \bibinfo {author} {\bibfnamefont {D.}~\bibnamefont {Rinberg}},\
  }\href@noop {} {\bibfield  {journal} {\bibinfo  {journal} {Nature
  neuroscience}\ }\textbf {\bibinfo {volume} {14}},\ \bibinfo {pages} {1039}
  (\bibinfo {year} {2011})}\BibitemShut {NoStop}%
\bibitem [{\citenamefont {Lawson}\ \emph {et~al.}(2012)\citenamefont {Lawson},
  \citenamefont {Craven}, \citenamefont {Paterson},\ and\ \citenamefont
  {Settles}}]{Craven2012}%
  \BibitemOpen
  \bibfield  {author} {\bibinfo {author} {\bibfnamefont {M.~J.}\ \bibnamefont
  {Lawson}}, \bibinfo {author} {\bibfnamefont {B.~A.}\ \bibnamefont {Craven}},
  \bibinfo {author} {\bibfnamefont {E.~G.}\ \bibnamefont {Paterson}}, \ and\
  \bibinfo {author} {\bibfnamefont {G.~S.}\ \bibnamefont {Settles}},\ }\href
  {\doibase 10.1093/chemse/bjs039} {\bibfield  {journal} {\bibinfo  {journal}
  {Chemical Senses}\ }\textbf {\bibinfo {volume} {37}},\ \bibinfo {pages} {553}
  (\bibinfo {year} {2012})},\ \Eprint
  {http://arxiv.org/abs/https://academic.oup.com/chemse/article-pdf/37/6/553/1037113/bjs039.pdf}
  {https://academic.oup.com/chemse/article-pdf/37/6/553/1037113/bjs039.pdf}
  \BibitemShut {NoStop}%
\bibitem [{\citenamefont {Kurtz}\ \emph {et~al.}(2004)\citenamefont {Kurtz},
  \citenamefont {Zhao}, \citenamefont {Hornung},\ and\ \citenamefont
  {Scherer}}]{Kurtzchemse2004}%
  \BibitemOpen
  \bibfield  {author} {\bibinfo {author} {\bibfnamefont {D.~B.}\ \bibnamefont
  {Kurtz}}, \bibinfo {author} {\bibfnamefont {K.}~\bibnamefont {Zhao}},
  \bibinfo {author} {\bibfnamefont {D.~E.}\ \bibnamefont {Hornung}}, \ and\
  \bibinfo {author} {\bibfnamefont {P.}~\bibnamefont {Scherer}},\ }\href
  {\doibase 10.1093/chemse/bjh079} {\bibfield  {journal} {\bibinfo  {journal}
  {Chemical Senses}\ }\textbf {\bibinfo {volume} {29}},\ \bibinfo {pages} {763}
  (\bibinfo {year} {2004})},\ \Eprint
  {http://arxiv.org/abs/https://academic.oup.com/chemse/article-pdf/29/9/763/758058/bjh079.pdf}
  {https://academic.oup.com/chemse/article-pdf/29/9/763/758058/bjh079.pdf}
  \BibitemShut {NoStop}%
\bibitem [{\citenamefont {Keyhani}\ \emph {et~al.}(1997)\citenamefont
  {Keyhani}, \citenamefont {Scherer},\ and\ \citenamefont
  {Mozell}}]{KEYHANI1997}%
  \BibitemOpen
  \bibfield  {author} {\bibinfo {author} {\bibfnamefont {K.}~\bibnamefont
  {Keyhani}}, \bibinfo {author} {\bibfnamefont {P.~W.}\ \bibnamefont
  {Scherer}}, \ and\ \bibinfo {author} {\bibfnamefont {M.~M.}\ \bibnamefont
  {Mozell}},\ }\href {\doibase https://doi.org/10.1006/jtbi.1996.0347}
  {\bibfield  {journal} {\bibinfo  {journal} {Journal of Theoretical Biology}\
  }\textbf {\bibinfo {volume} {186}},\ \bibinfo {pages} {279} (\bibinfo {year}
  {1997})}\BibitemShut {NoStop}%
\bibitem [{\citenamefont {Tritton}(2012)}]{Tritton}%
  \BibitemOpen
  \bibfield  {author} {\bibinfo {author} {\bibfnamefont {D.~J.}\ \bibnamefont
  {Tritton}},\ }\href@noop {} {\emph {\bibinfo {title} {Physical fluid
  dynamics}}}\ (\bibinfo  {publisher} {Springer Science \& Business Media},\
  \bibinfo {year} {2012})\BibitemShut {NoStop}%
\bibitem [{\citenamefont {Shang}\ \emph {et~al.}(2021)\citenamefont {Shang},
  \citenamefont {Inthavong}, \citenamefont {Qiu}, \citenamefont {Singh},
  \citenamefont {He},\ and\ \citenamefont {Tu}}]{shangPLoSOne2021}%
  \BibitemOpen
  \bibfield  {author} {\bibinfo {author} {\bibfnamefont {Y.}~\bibnamefont
  {Shang}}, \bibinfo {author} {\bibfnamefont {K.}~\bibnamefont {Inthavong}},
  \bibinfo {author} {\bibfnamefont {D.}~\bibnamefont {Qiu}}, \bibinfo {author}
  {\bibfnamefont {N.}~\bibnamefont {Singh}}, \bibinfo {author} {\bibfnamefont
  {F.}~\bibnamefont {He}}, \ and\ \bibinfo {author} {\bibfnamefont
  {J.}~\bibnamefont {Tu}},\ }\href@noop {} {\bibfield  {journal} {\bibinfo
  {journal} {PLoS One}\ }\textbf {\bibinfo {volume} {16}},\ \bibinfo {pages}
  {e0246007} (\bibinfo {year} {2021})}\BibitemShut {NoStop}%
\bibitem [{\citenamefont {Quraishi}\ \emph {et~al.}(1998)\citenamefont
  {Quraishi}, \citenamefont {Jones},\ and\ \citenamefont
  {Mason}}]{Quraishi1998}%
  \BibitemOpen
  \bibfield  {author} {\bibinfo {author} {\bibfnamefont {M.~S.}\ \bibnamefont
  {Quraishi}}, \bibinfo {author} {\bibfnamefont {N.~S.}\ \bibnamefont {Jones}},
  \ and\ \bibinfo {author} {\bibfnamefont {J.}~\bibnamefont {Mason}},\ }\href
  {\doibase https://doi.org/10.1046/j.1365-2273.1998.00172.x} {\bibfield
  {journal} {\bibinfo  {journal} {Clinical Otolaryngology \& Allied Sciences}\
  }\textbf {\bibinfo {volume} {23}},\ \bibinfo {pages} {403} (\bibinfo {year}
  {1998})},\ \Eprint
  {http://arxiv.org/abs/https://onlinelibrary.wiley.com/doi/pdf/10.1046/j.1365-2273.1998.00172.x}
  {https://onlinelibrary.wiley.com/doi/pdf/10.1046/j.1365-2273.1998.00172.x}
  \BibitemShut {NoStop}%
\bibitem [{Beu(2010)}]{Beule2010}%
  \BibitemOpen
  \href {\doibase 10.3205/cto000071} {\  (\bibinfo {year} {2010}),\
  10.3205/cto000071}\BibitemShut {NoStop}%
\bibitem [{\citenamefont {Sun}\ \emph {et~al.}(2018)\citenamefont {Sun},
  \citenamefont {Xiao},\ and\ \citenamefont {Carlson}}]{Jennifer2018}%
  \BibitemOpen
  \bibfield  {author} {\bibinfo {author} {\bibfnamefont {J.~S.}\ \bibnamefont
  {Sun}}, \bibinfo {author} {\bibfnamefont {S.}~\bibnamefont {Xiao}}, \ and\
  \bibinfo {author} {\bibfnamefont {J.~R.}\ \bibnamefont {Carlson}},\ }\href
  {\doibase 10.1098/rsob.180208} {\bibfield  {journal} {\bibinfo  {journal}
  {Open Biology}\ }\textbf {\bibinfo {volume} {8}},\ \bibinfo {pages} {180208}
  (\bibinfo {year} {2018})},\ \Eprint
  {http://arxiv.org/abs/https://royalsocietypublishing.org/doi/pdf/10.1098/rsob.180208}
  {https://royalsocietypublishing.org/doi/pdf/10.1098/rsob.180208} \BibitemShut
  {NoStop}%
\bibitem [{\citenamefont {Brito}\ \emph {et~al.}(2016)\citenamefont {Brito},
  \citenamefont {Moreira},\ and\ \citenamefont {Melo}}]{Brito2016}%
  \BibitemOpen
  \bibfield  {author} {\bibinfo {author} {\bibfnamefont {N.~F.}\ \bibnamefont
  {Brito}}, \bibinfo {author} {\bibfnamefont {M.~F.}\ \bibnamefont {Moreira}},
  \ and\ \bibinfo {author} {\bibfnamefont {A.~C.}\ \bibnamefont {Melo}},\
  }\href {\doibase https://doi.org/10.1016/j.jinsphys.2016.09.008} {\bibfield
  {journal} {\bibinfo  {journal} {Journal of Insect Physiology}\ }\textbf
  {\bibinfo {volume} {95}},\ \bibinfo {pages} {51} (\bibinfo {year}
  {2016})}\BibitemShut {NoStop}%
\bibitem [{\citenamefont {Larter}\ \emph {et~al.}(2016)\citenamefont {Larter},
  \citenamefont {Sun},\ and\ \citenamefont {Carlson}}]{Larter2016}%
  \BibitemOpen
  \bibfield  {author} {\bibinfo {author} {\bibfnamefont {N.~K.}\ \bibnamefont
  {Larter}}, \bibinfo {author} {\bibfnamefont {J.~S.}\ \bibnamefont {Sun}}, \
  and\ \bibinfo {author} {\bibfnamefont {J.~R.}\ \bibnamefont {Carlson}},\
  }\href {\doibase 10.7554/eLife.20242} {\bibfield  {journal} {\bibinfo
  {journal} {eLife}\ }\textbf {\bibinfo {volume} {5}},\ \bibinfo {pages}
  {e20242} (\bibinfo {year} {2016})}\BibitemShut {NoStop}%
\bibitem [{\citenamefont {Strotmann}\ and\ \citenamefont
  {Breer}(2011)}]{strotmann2011}%
  \BibitemOpen
  \bibfield  {author} {\bibinfo {author} {\bibfnamefont {J.}~\bibnamefont
  {Strotmann}}\ and\ \bibinfo {author} {\bibfnamefont {H.}~\bibnamefont
  {Breer}},\ }\href@noop {} {\bibfield  {journal} {\bibinfo  {journal}
  {Histochemistry and cell biology}\ }\textbf {\bibinfo {volume} {136}},\
  \bibinfo {pages} {357} (\bibinfo {year} {2011})}\BibitemShut {NoStop}%
\bibitem [{\citenamefont {Cover}\ and\ \citenamefont
  {Thomas}(1991)}]{ITbook1991}%
  \BibitemOpen
  \bibfield  {author} {\bibinfo {author} {\bibfnamefont {T.}~\bibnamefont
  {Cover}}\ and\ \bibinfo {author} {\bibfnamefont {J.}~\bibnamefont {Thomas}},\
  }\href {https://books.google.de/books?id=CX9QAAAAMAAJ} {\emph {\bibinfo
  {title} {Elements of Information Theory}}},\ Wiley Series in
  Telecommunications and Signal Processing\ (\bibinfo  {publisher} {Wiley},\
  \bibinfo {year} {1991})\BibitemShut {NoStop}%
\bibitem [{\citenamefont {Zwicker}(2016)}]{zwickerPlosone2016}%
  \BibitemOpen
  \bibfield  {author} {\bibinfo {author} {\bibfnamefont {D.}~\bibnamefont
  {Zwicker}},\ }\href@noop {} {\bibfield  {journal} {\bibinfo  {journal} {PLoS
  One}\ }\textbf {\bibinfo {volume} {11}},\ \bibinfo {pages} {e0166456}
  (\bibinfo {year} {2016})}\BibitemShut {NoStop}%
\bibitem [{\citenamefont {Zwicker}(2019)}]{zwickerPloscompbio2019}%
  \BibitemOpen
  \bibfield  {author} {\bibinfo {author} {\bibfnamefont {D.}~\bibnamefont
  {Zwicker}},\ }\href@noop {} {\bibfield  {journal} {\bibinfo  {journal} {PLoS
  Computational Biology}\ }\textbf {\bibinfo {volume} {15}},\ \bibinfo {pages}
  {e1007188} (\bibinfo {year} {2019})}\BibitemShut {NoStop}%
\bibitem [{\citenamefont {Challis}\ \emph {et~al.}(2015)\citenamefont
  {Challis}, \citenamefont {Tian}, \citenamefont {Wang}, \citenamefont {He},
  \citenamefont {Jiang}, \citenamefont {Chen}, \citenamefont {Yin},
  \citenamefont {Connelly}, \citenamefont {Ma}, \citenamefont {Yu},
  \citenamefont {Pluznick}, \citenamefont {Storm}, \citenamefont {Huang},
  \citenamefont {Zhao},\ and\ \citenamefont {Ma}}]{Rosemary2015}%
  \BibitemOpen
  \bibfield  {author} {\bibinfo {author} {\bibfnamefont {R.}~\bibnamefont
  {Challis}}, \bibinfo {author} {\bibfnamefont {H.}~\bibnamefont {Tian}},
  \bibinfo {author} {\bibfnamefont {J.}~\bibnamefont {Wang}}, \bibinfo {author}
  {\bibfnamefont {J.}~\bibnamefont {He}}, \bibinfo {author} {\bibfnamefont
  {J.}~\bibnamefont {Jiang}}, \bibinfo {author} {\bibfnamefont
  {X.}~\bibnamefont {Chen}}, \bibinfo {author} {\bibfnamefont {W.}~\bibnamefont
  {Yin}}, \bibinfo {author} {\bibfnamefont {T.}~\bibnamefont {Connelly}},
  \bibinfo {author} {\bibfnamefont {L.}~\bibnamefont {Ma}}, \bibinfo {author}
  {\bibfnamefont {C.}~\bibnamefont {Yu}}, \bibinfo {author} {\bibfnamefont
  {J.}~\bibnamefont {Pluznick}}, \bibinfo {author} {\bibfnamefont
  {D.}~\bibnamefont {Storm}}, \bibinfo {author} {\bibfnamefont
  {L.}~\bibnamefont {Huang}}, \bibinfo {author} {\bibfnamefont
  {K.}~\bibnamefont {Zhao}}, \ and\ \bibinfo {author} {\bibfnamefont
  {M.}~\bibnamefont {Ma}},\ }\href {\doibase
  https://doi.org/10.1016/j.cub.2015.07.065} {\bibfield  {journal} {\bibinfo
  {journal} {Current Biology}\ }\textbf {\bibinfo {volume} {25}},\ \bibinfo
  {pages} {2503} (\bibinfo {year} {2015})}\BibitemShut {NoStop}%
\bibitem [{\citenamefont {Ortega}\ \emph {et~al.}(2020)\citenamefont {Ortega},
  \citenamefont {Mariottini}, \citenamefont {Troina}, \citenamefont
  {Dahlquist}, \citenamefont {Ricci},\ and\ \citenamefont
  {Plaxco}}]{Gabriel2020}%
  \BibitemOpen
  \bibfield  {author} {\bibinfo {author} {\bibfnamefont {G.}~\bibnamefont
  {Ortega}}, \bibinfo {author} {\bibfnamefont {D.}~\bibnamefont {Mariottini}},
  \bibinfo {author} {\bibfnamefont {A.}~\bibnamefont {Troina}}, \bibinfo
  {author} {\bibfnamefont {F.~W.}\ \bibnamefont {Dahlquist}}, \bibinfo {author}
  {\bibfnamefont {F.}~\bibnamefont {Ricci}}, \ and\ \bibinfo {author}
  {\bibfnamefont {K.~W.}\ \bibnamefont {Plaxco}},\ }\href {\doibase
  10.1073/pnas.2006254117} {\bibfield  {journal} {\bibinfo  {journal}
  {Proceedings of the National Academy of Sciences}\ }\textbf {\bibinfo
  {volume} {117}},\ \bibinfo {pages} {19136} (\bibinfo {year} {2020})},\
  \Eprint
  {http://arxiv.org/abs/https://www.pnas.org/doi/pdf/10.1073/pnas.2006254117}
  {https://www.pnas.org/doi/pdf/10.1073/pnas.2006254117} \BibitemShut {NoStop}%
\bibitem [{\citenamefont {Singh}\ \emph {et~al.}(2019)\citenamefont {Singh},
  \citenamefont {Murphy}, \citenamefont {Balasubramanian},\ and\ \citenamefont
  {Mainland}}]{BalasubramanianPNAS2019}%
  \BibitemOpen
  \bibfield  {author} {\bibinfo {author} {\bibfnamefont {V.}~\bibnamefont
  {Singh}}, \bibinfo {author} {\bibfnamefont {N.~R.}\ \bibnamefont {Murphy}},
  \bibinfo {author} {\bibfnamefont {V.}~\bibnamefont {Balasubramanian}}, \ and\
  \bibinfo {author} {\bibfnamefont {J.~D.}\ \bibnamefont {Mainland}},\ }\href
  {\doibase 10.1073/pnas.1813230116} {\bibfield  {journal} {\bibinfo  {journal}
  {Proceedings of the National Academy of Sciences}\ }\textbf {\bibinfo
  {volume} {116}},\ \bibinfo {pages} {9598} (\bibinfo {year} {2019})},\ \Eprint
  {http://arxiv.org/abs/https://www.pnas.org/doi/pdf/10.1073/pnas.1813230116}
  {https://www.pnas.org/doi/pdf/10.1073/pnas.1813230116} \BibitemShut {NoStop}%
\bibitem [{\citenamefont {Qin}\ \emph {et~al.}(2019)\citenamefont {Qin},
  \citenamefont {Li}, \citenamefont {Tang},\ and\ \citenamefont
  {Tu}}]{Shanshan2019}%
  \BibitemOpen
  \bibfield  {author} {\bibinfo {author} {\bibfnamefont {S.}~\bibnamefont
  {Qin}}, \bibinfo {author} {\bibfnamefont {Q.}~\bibnamefont {Li}}, \bibinfo
  {author} {\bibfnamefont {C.}~\bibnamefont {Tang}}, \ and\ \bibinfo {author}
  {\bibfnamefont {Y.}~\bibnamefont {Tu}},\ }\href {\doibase
  10.1073/pnas.1906571116} {\bibfield  {journal} {\bibinfo  {journal}
  {Proceedings of the National Academy of Sciences}\ }\textbf {\bibinfo
  {volume} {116}},\ \bibinfo {pages} {20286} (\bibinfo {year} {2019})},\
  \Eprint
  {http://arxiv.org/abs/https://www.pnas.org/doi/pdf/10.1073/pnas.1906571116}
  {https://www.pnas.org/doi/pdf/10.1073/pnas.1906571116} \BibitemShut {NoStop}%
\bibitem [{\citenamefont {Reddy}\ \emph {et~al.}(2018)\citenamefont {Reddy},
  \citenamefont {Zak}, \citenamefont {Vergassola},\ and\ \citenamefont
  {Murthy}}]{Reddy2018}%
  \BibitemOpen
  \bibfield  {author} {\bibinfo {author} {\bibfnamefont {G.}~\bibnamefont
  {Reddy}}, \bibinfo {author} {\bibfnamefont {J.~D.}\ \bibnamefont {Zak}},
  \bibinfo {author} {\bibfnamefont {M.}~\bibnamefont {Vergassola}}, \ and\
  \bibinfo {author} {\bibfnamefont {V.~N.}\ \bibnamefont {Murthy}},\ }\href
  {\doibase 10.7554/eLife.34958} {\bibfield  {journal} {\bibinfo  {journal}
  {eLife}\ }\textbf {\bibinfo {volume} {7}},\ \bibinfo {pages} {e34958}
  (\bibinfo {year} {2018})}\BibitemShut {NoStop}%
\bibitem [{\citenamefont {Coleman}\ \emph {et~al.}(2019)\citenamefont
  {Coleman}, \citenamefont {Lin}, \citenamefont {Louie}, \citenamefont
  {Peterson}, \citenamefont {Lane},\ and\ \citenamefont
  {Schwob}}]{colemanJNeu2019}%
  \BibitemOpen
  \bibfield  {author} {\bibinfo {author} {\bibfnamefont {J.~H.}\ \bibnamefont
  {Coleman}}, \bibinfo {author} {\bibfnamefont {B.}~\bibnamefont {Lin}},
  \bibinfo {author} {\bibfnamefont {J.~D.}\ \bibnamefont {Louie}}, \bibinfo
  {author} {\bibfnamefont {J.}~\bibnamefont {Peterson}}, \bibinfo {author}
  {\bibfnamefont {R.~P.}\ \bibnamefont {Lane}}, \ and\ \bibinfo {author}
  {\bibfnamefont {J.~E.}\ \bibnamefont {Schwob}},\ }\href@noop {} {\bibfield
  {journal} {\bibinfo  {journal} {Journal of Neuroscience}\ }\textbf {\bibinfo
  {volume} {39}},\ \bibinfo {pages} {814} (\bibinfo {year} {2019})}\BibitemShut
  {NoStop}%
\bibitem [{\citenamefont {Horgue}\ \emph {et~al.}(2022)\citenamefont {Horgue},
  \citenamefont {Assens}, \citenamefont {Fodoulian}, \citenamefont {Marconi},
  \citenamefont {Tuberosa}, \citenamefont {Haider}, \citenamefont {Boillat},
  \citenamefont {Carleton},\ and\ \citenamefont {Rodriguez}}]{horgue2022}%
  \BibitemOpen
  \bibfield  {author} {\bibinfo {author} {\bibfnamefont {L.~F.}\ \bibnamefont
  {Horgue}}, \bibinfo {author} {\bibfnamefont {A.}~\bibnamefont {Assens}},
  \bibinfo {author} {\bibfnamefont {L.}~\bibnamefont {Fodoulian}}, \bibinfo
  {author} {\bibfnamefont {L.}~\bibnamefont {Marconi}}, \bibinfo {author}
  {\bibfnamefont {J.}~\bibnamefont {Tuberosa}}, \bibinfo {author}
  {\bibfnamefont {A.}~\bibnamefont {Haider}}, \bibinfo {author} {\bibfnamefont
  {M.}~\bibnamefont {Boillat}}, \bibinfo {author} {\bibfnamefont
  {A.}~\bibnamefont {Carleton}}, \ and\ \bibinfo {author} {\bibfnamefont
  {I.}~\bibnamefont {Rodriguez}},\ }\href@noop {} {\bibfield  {journal}
  {\bibinfo  {journal} {Nature communications}\ }\textbf {\bibinfo {volume}
  {13}},\ \bibinfo {pages} {1} (\bibinfo {year} {2022})}\BibitemShut {NoStop}%
\bibitem [{\citenamefont {Raman}\ \emph {et~al.}(2011)\citenamefont {Raman},
  \citenamefont {Stopfer},\ and\ \citenamefont {Semancik}}]{Raman2011}%
  \BibitemOpen
  \bibfield  {author} {\bibinfo {author} {\bibfnamefont {B.}~\bibnamefont
  {Raman}}, \bibinfo {author} {\bibfnamefont {M.}~\bibnamefont {Stopfer}}, \
  and\ \bibinfo {author} {\bibfnamefont {S.}~\bibnamefont {Semancik}},\ }\href
  {\doibase 10.1021/cn200027r} {\bibfield  {journal} {\bibinfo  {journal} {ACS
  Chemical Neuroscience}\ }\textbf {\bibinfo {volume} {2}},\ \bibinfo {pages}
  {487} (\bibinfo {year} {2011})},\ \bibinfo {note} {pMID: 22081790},\ \Eprint
  {http://arxiv.org/abs/https://doi.org/10.1021/cn200027r}
  {https://doi.org/10.1021/cn200027r} \BibitemShut {NoStop}%
\bibitem [{\citenamefont {Tong}(1990)}]{tong1990multivariate}%
  \BibitemOpen
  \bibfield  {author} {\bibinfo {author} {\bibfnamefont {Y.}~\bibnamefont
  {Tong}},\ }\href {https://books.google.de/books?id=H\_q3vORwyJcC} {\emph
  {\bibinfo {title} {The Multivariate Normal Distribution}}},\ Research in
  Criminology\ (\bibinfo  {publisher} {Springer New York},\ \bibinfo {year}
  {1990})\BibitemShut {NoStop}%
\bibitem [{\citenamefont {Forney}\ and\ \citenamefont
  {Ungerboeck}(1998)}]{Forney1998}%
  \BibitemOpen
  \bibfield  {author} {\bibinfo {author} {\bibfnamefont {G.}~\bibnamefont
  {Forney}}\ and\ \bibinfo {author} {\bibfnamefont {G.}~\bibnamefont
  {Ungerboeck}},\ }\href {\doibase 10.1109/18.720542} {\bibfield  {journal}
  {\bibinfo  {journal} {IEEE Transactions on Information Theory}\ }\textbf
  {\bibinfo {volume} {44}},\ \bibinfo {pages} {2384} (\bibinfo {year}
  {1998})}\BibitemShut {NoStop}%
\end{thebibliography}%

\end{document}